\documentclass{aa}  

\usepackage{graphicx}
\usepackage{txfonts}
\usepackage{multicol}
\usepackage{url}
\usepackage{longtable}
\usepackage{subcaption}         
\usepackage{lscape}             
\usepackage{placeins}           
                                
\begin{document}

   \title{OH/IR stars in the inner Galactic bulge}
   \subtitle{I. Colour, CO line, and stellar light curve characteristics}
   

   \author{H.~Olofsson   \inst{1}
          \and
          T.~Khouri            \inst{1}
          \and
          S.~Muller              \inst{1}
          \and
          R.~Unnikrishnan   \inst{1}
          \and
          M.A.T.~Groenewegen \inst{2}
          \and
          J.A.D.L. Blommaert \inst{3}
          \and
          E.~De Beck           \inst{1} 
          \and
          J.H.~Kastner \inst{4}
          \and
          M. Maercker \inst{1}
          \and
          N.~Patel \inst{5}
          \and
          N.~Ryde \inst{6}
          \and
          B.A.~Sargent  \inst{7,8}
          \and
          S.~Srinivasan \inst{9}
          \and
          W.H.T.~Vlemmings     \inst{1}
          }

\institute{Department of Physics and Astronomy, Chalmers University of Technology, SE-41296 G{\"o}teborg, Sweden\\
         \email{hans.olofsson@chalmers.se}
         \and
         Koninklijke Sterrenwacht van Belgi\"e, Ringlaan 3, B--1180 Brussels, Belgium
         \and
         Astronomy and Astrophysics Research Group, Department of Physics and Astrophysics, Vrije Universiteit Brussel, Pleinlaan 2, B-1050 Brussels, Belgium
         \and
         Chester F. Carlson Center for Imaging Science, School of Physics \& Astronomy, and Laboratory for Multiwavelength Astrophysics, Rochester Institute of Technology, Rochester, NY 14623, USA
         \and
         Center for Astrophysics | Harvard \& Smithsonian, 60 Garden Street, MS 78, Cambridge, MA 02138, USA
         \and
         Division of Astrophysics, Department of Physics, Lund University, Box 118, SE-221 00 Lund, Sweden
         \and
         Space Telescope Science Institute, 3700 San Martin Drive, Baltimore, MD 21218, USA
         \and
         Center for Astrophysical Sciences, The William H. Miller III Department of Physics and Astronomy, Johns Hopkins University, Baltimore, MD 21218, USA
         \and
         Instituto de Radioastronom\'ia y Astrof\'isica, Universidad Nacional Aut\'onoma de M\'exico, Morelia, Michoac\'an, 58089 M\'exico              
}
   \date{Received 9 April 2026 / Accepted 11 May 2026}


 \abstract
   {Stars on the asymptotic giant branch (AGB) play important roles in a number of astronomical contexts. 
   To quantify these roles, it is necessary to establish the mass-loss characteristics of stars as they evolve up and beyond the AGB.}
   {We used an equidistant sample of 77 AGB stars in the inner Galactic bulge, selected on the existence and strength of OH1612\,MHz maser emission, to obtain information on the mass loss of O-rich AGB stars, and on its transformations in geometry and kinematics that occur at the tip of the AGB.} 
   {Observations of circumstellar lines from several rotational transitions of  $^{12}$CO, $^{13}$CO, and C$^{18}$O were performed with the Atacama Large Millimeter/submillimeter Array (ALMA), and, for a sub-sample, with the Atacama Pathfinder EXperiment telescope (APEX). The ALMA observations also provide continuum data. Existing infrared photometry was used to estimate colours and produce stellar light curves.}
   {Based on mid-infrared colour, CO line, and near-infrared variability characteristics, the objects were divided into four categories of distinct stellar and/or circumstellar properties. Various circumstellar CO line characteristics are presented and compared between the categories, such as morphologies and extents of brightness distributions (BDs), line profiles (LPs), line intensity ratios, and kinematics. A majority of the objects form a homogenous group with sharply, centrally peaked BDs and LPs of the soft-parabola type, while the rest show extended and complex BDs and/or LPs with high-velocity wings. The C$^{18}$O line and ALMA continuum detection rates vary significantly between the categories. CO line emission is also detected from interaction zones where the expanding circumstellar gas meets the interstellar medium.}
   {These data provide the foundation for more detailed studies, including radiative transfer analyses of the CO line and photometry data, on the evolution of stellar and circumstellar characteristics of O-rich stars on the upper AGB and beyond. }


   \maketitle
\nolinenumbers

\section{Introduction}

Low- to intermediate-mass stars ($\approx$\,1--8\,$M_\odot$) end their stellar evolution on the asymptotic giant branch \citep[AGB; ][]{herw05}. Taking into account the stellar initial mass function, stellar lifetimes as a function of their mass, and the age of the Universe, it is clear that the majority of stars that have died in our Universe did so following the AGB phase. Hence, a full understanding of the evolution during this phase is of utmost importance for establishing the way AGB stars contribute in more general contexts, such as in galactic chemical evolution
(\citep{kara16}, including dust production, where the relative contributions of supernovae and dust growth in the ISM is an ongoing discussion \citep{zhuketal18}, galaxy luminosities at infrared wavelengths \citep[e.g.][]{villetal15}, and the relation between AGB stars and planetary nebulae \citep[PNe;][]{kwok24}.

Stellar evolution on the AGB is to a large extent driven by mass loss from the surface \citep{hoefolof18}. It is therefore important to observationally constrain its characteristics (such as its magnitude and geometry) on the AGB and beyond. For this, a well-defined sample is required. It has recently been shown that a sample of equidistant Galactic AGB stars, detectable in circumstellar CO lines with the Atacama Large Millimeter/submillimeter Array (ALMA), can be obtained through selecting OH/IR stars located in the inner Galactic bulge  \citep[][hereafter O+22]{olofetal22}. All AGB stars for which oxygen is more common than carbon by number (often termed O-rich) are believed to become OH/IR stars when their mass-loss rates (MLRs) are high enough, a few 10$^{-6}$\,$M_\odot$\,yr$^{-1}$, to produce a strong OH 1612\,MHz maser  \citep{elitetal76, habi96}. They are easily identified through the double-peaked maser line profile.  OH/IR stars have thick circumstellar envelopes (CSEs) and reach MLRs well in excess of 10$^{-5}$\,$M_\odot$\,yr$^{-1}$, with the extreme ones in terms of MLR expected to be located close to the tip of the AGB. Some may even have stopped losing mass and evolved off the AGB, but only very recently.  The presence of OH maser emission from an object puts a limit of, at most, about a thousand years since the cessation of its mass loss \citep{engeetal19}. These are the stars that are believed to contribute significantly to the return of gas (including newly synthesised elements) and dust to the ISM by low- to intermediate-mass stars. Among these, one also finds a large fraction of possible PN precursors. Considering the limited information on mass-loss characteristics extractable from OH maser emission, a detailed study of OH/IR stars based on observations of CO rotational lines is warranted.

The OH/IR sample of O+22, 22 objects within 2$^\circ$ of the Galactic centre, was observed in $^{12}$CO and $^{13}$CO lines using ALMA. It was concluded that 15 are located at the distance of the Galactic centre, and that they are of lower mass (median mass $\approx$\,1.4\,$M_\odot$, with a population age of $\approx$\,4$-$7\,Gyr), have a median luminosity of 5400\,$L_\odot$, and a median MLR of 2$\times$10$^{-5}$\,$M_\odot$\,yr$^{-1}$, as estimated using a radiative transfer model for the CO line emission. These inner Galactic bulge OH/IR stars appear to have the same characteristics as Galactic disk OH/IR stars \citep{habi96}.

The study of O+22 had limitations. Only two $^{12}$CO lines and one $^{13}$CO line were observed, often with limited S/N.  The limited sample size (15 objects) prevented statistical comparisons. In the present study we remedy these shortcomings by observing 77 objects in the inner Galactic bulge in different CO isotopologue lines with ALMA and a sub-sample with the Atacama Pathfinder EXperiment telescope (APEX). and by determining their spectral energy distributions (SEDs) and variability characteristics. This study of an equidistant sample will complement other studies of the circumstellar properties of AGB stars, such as the ATOMIUM \citep{decietal20}, DEATHSTAR  \citep{ramsetal20, andretal22}, and NESS projects \citep{scicetal22, walletal25}. 

We describe the sample selection in Sect.~\ref{s:sample}. The ALMA and APEX observations and data reduction, as well as the photometric data, required for colour-colour diagrams and variability analysis, are presented in Sect.~\ref{s:observations}. A division of our objects into categories is introduced in Sect.~\ref{s:categories}. The ALMA results in terms of CO line brightness distributions, flux densities, and line shapes, and continuum flux densities are presented in Sects~\ref{s:results_abc} and \ref{s:results_d}, and a comparison between ALMA and APEX CO line intensities is given in Sect.~\ref{s:alma_apex_co}.  The results are summarised in Sect.~\ref{s:summary}.
%
%
%
%
\section{Sample}
\label{s:sample}

OH 1612\,MHz maser emission has been important for studies of evolved stars throughout the Galaxy \citep{chenetal01, engebunz15}.  \citet{vandvehabi90} and \citet{blometal18} concluded that the OH/IR stars in the Galactic bulge are dominated by $\approx$\,1.0--1.5\,$M_\odot$ progenitor stars of age 3\,$-$\,7\,Gyr. \citet{woodetal98} studied OH/IR stars within 0.7$^\circ$ of the Galactic centre and concluded that closer to the centre there are also younger AGB stars belonging to an intermediate-age population ($\approx$\,0.5\,Gyr) with masses above 4\,$M_\odot$. 

The survey of OH/IR stars by \citet{seveetal97a} forms the basis of our sample. It is estimated to be complete in the region 0.3$^\circ$ to 3$^\circ$ of the Galactic centre (the inner Galactic bulge; the very-high-extinction area centred on the Galactic centre is excluded) down to an OH 1612\,MHz flux density of 0.3\,Jy. Thus, our objects are expected to represent the high-MLR phase of O-rich stars on and (slightly) beyond the AGB. We selected all objects, brighter than 0.3\,Jy in the OH 1612\,MHz line, that have an estimated luminosity within a factor of two of the median luminosity of the O+22 sample, 5400\,$L_\odot$. This minimized contamination by foreground objects, and helped ensure that the sample covers a narrow range in initial mass. Based on the results of O+22, a flux density of 0.3\,Jy corresponds to a MLR of $\approx$\,5$\times$10$^{-6}$\,$M_\odot$\,yr$^{-1}$. This is a reasonable lower limit, since below this value not all O-rich AGB stars develop strong OH 1612\,MHz maser emission. The final sample consists of 77 objects. They are shown overlaid on an image of interstellar CO line emission in Fig.~\ref{f:sample_co}, and listed in Table~\ref{t:sample}. The adopted distance to all of our sources is 8.2\,kpc \citep{gravcoll19}, even though some spread along the line of sight is most likely. Unfortunately, none of our sources have a measured reliable parallax in Gaia Early Data Release 3 \citep{gaiacoll23}. 

%
%
%
%
\section{Observations and data reduction}
\label{s:observations}

\subsection{ALMA data}
\label{s:alma}

The sample of 77 OH/IR stars was observed with ALMA band 3 \citep[B3; ][]{clauetal08}, 6 \citep[B6; ][]{edisetal04}, and 7 \citep[B7; ][]{mahietal12} receivers to cover the $^{12}$CO $J$\,=\,\mbox{1--0}, \mbox{2--1}, and \mbox{3--2}, and the $^{13}$CO $J$\,=\,\mbox{2--1} and \mbox{3--2} lines (project 2023.1.00098.S). The B6 setting also covered the C$^{18}$O $J$\,=\,\mbox{2--1} line. All observations were done in July 2024 in the C6 configuration (baselines between $\approx$\,15 and 2500\,m), except one visit in B3 in January 2024; this observation was done with the more compact C4 configuration (baselines between $\approx$\,15 and 800\,m). The resulting synthesised beams are between $\approx$\,0\farcs15\,--\,0\farcs4 (from B7 to B3). The maximum recoverable scales (MRSs) are about 5\farcs2, 2\farcs4, and 1\farcs9 in B3, B6, and B7, respectively. A journal of the observations is given in Table~\ref{t:journal}. Since all the targets are within a region less than 6$^\circ$ in extent, the observations were done in a snapshot mode, with an integration time of half a minute per star per visit, and with all stars sharing the same calibration. The radio-frequency bandpass response of the antennas was calibrated with the bright quasar J\,1924$-$2914. The gain calibration (amplitude and phase) was performed on J\,1744$-$3116. The flux scale was calibrated using the quasar J\,1924$-$2914 for all observations except for the first visit of the B3 tuning (January 2024), when J\,1517$-$2422 was used. The data calibration was done with the standard pipeline procedure under the Common Astronomy Software Applications (CASA) package, version 6.5.4.9 \citep{mcmuetal07}.\footnote{\tt http://casa.nrao.edu/} The calibrators are quite stable and monitored often.\footnote{\url{https://almascience.eso.org/alma-data/calibrator-catalogue}} Based on the scatter and time series data for the calibrators, the absolute flux uncertainty is estimated to be $\la$\,5\,\% in all three bands.

In B3, the correlator was set up with four spectral windows centred at the sky frequencies 100.863, 102.673, 112.865, and 114.756\,GHz ($^{12}$CO(\mbox{1--0})). Each spectral window had a width of 1.875\,GHz covered by 1920 channels. In B6, four spectral windows were centred at sky frequencies 218.944 (C$^{18}$O(\mbox{2--1})), 220.340 ($^{13}$CO(\mbox{2--1})), 230.476 ($^{12}$CO(\mbox{2--1})), and 232.944\,GHz. All windows were covered with 960 channels. The windows centred on the $^{12}$CO and $^{13}$CO lines had a width of 938\,MHz, while the two others had a width of 1.875\,GHz. Due to the large range in source systemic velocities (from $-$340 to +307\,km\,s$^{-1}$), the B7 observations were divided into two sub-samples, one for sources with negative line-of-sight velocities centred at 331.091 ($^{13}$CO(\mbox{3--2})), 332.595, 344.595, and 345.895\,GHz ($^{12}$CO(\mbox{3--2})) and the other for sources with positive line-of-sight velocities centred at 330.540 ($^{13}$CO(\mbox{3--2})), 331.940, 344.085, and 345.485\,GHz ($^{12}$CO(\mbox{3--2})), in order to simultaneously observe the $^{12}$CO and $^{13}$CO \mbox{3--2} lines for all sources. Despite this, part of the $^{13}$CO(\mbox{3--2}) line was missed for seven objects (for three of these objects the data of O+22 were used instead). Only three objects were not detected in any CO line.

\begin{table}
\caption{Journal of the observations.}
\centering
\label{t:journal}
\begin{tabular}{lcccc}
\hline \hline
\\[-2ex] 
Band                     & Date of              & $N_{\rm ant}$\,$^a$      & PWV\,$^b$          &  $\theta_{\rm beam}$\,$^c$\\
                             & observation       &                                            & [mm]                         &  [\arcsec $\times$ \arcsec, PA ($^\circ$)] \\
 \hline
 \\[-2ex] 
B3                         & 2024 Jan 05      & 32                                       & 5.9                            & 0.45 x 0.39, $-73$ \\
                             & 2024 Jul 03       & 44                                       & 0.3  \\
                             & 2024 Jul 09       & 42                                       & 0.9  \\
B6                        & 2024 Jul 14       & 43                                        & 0.4                            & 0.25 x 0.14, $-78$ \\
                            & 2024 Jul 14       & 42                                        & 0.4  \\
                            & 2024 Jul 17       & 43                                        & 1.0  \\
B7$-$\,$^d$         & 2024 Jul 02       & 43                                        & 0.2                           & 0.18 x 0.10, $-72$ \\
                            & 2024 Jul 18       & 46                                        & 0.6 & \\
B7$+$\,$^e$        & 2024 Jul 12       & 37                                        & 0.3                           & 0.18 x 0.10, $-72$ \\
                            & 2024 Jul 13       & 40                                        & 0.2  \\
 \hline
\end{tabular}
\tablefoot{$^{(a)}$ Number of 12\,m antennas in the array. $^{(b)}$ Amount of precipitable water vapour in the atmosphere. $^{(c)}$ Resulting synthesised beam (the numbers are indicative since the beam size differs somewhat from object to object). $^{(d)}$ Tuning for sources with negative line-of-sight velocities. $^{(e)}$ Tuning for sources with positive line-of-sight velocities.}
\end{table}

Imaging was done using the CASA {\tt tclean} algorithm with Briggs weighting (robust parameter set to 0.5) and primary beam correction was applied. The resulting beam sizes are given in Table~\ref{t:journal}. For the final line analysis the velocity resolutions at B3, B6, and B7 were binned to 3, 2, and 2\,km\,s$^{-1}$, respectively. The continuum images were produced using as much as possible of the line-free frequency ranges. Within each band the four different sub-bands were added to give the effective frequencies 107, 233, and 338\,GHz. The rms noise values per channel in the final line images are about 7, 4, and 5\,mJy\,beam$^{-1}$ in B3, B6, and B7, respectively. The continuum rms values are about 0.06, 0.1, and 0.2\,mJy\,beam$^{-1}$ at 107, 233, and 338\,GHz, respectively.

The source positions, both in lines and continuum, were determined by fitting two-dimensional Gaussians to the images (in the line case using the moment 0 images integrated over the velocity range of the circumstellar line). We estimate the astrometric accuracy to be better than 0\farcs1. The source positions, i.e. the peaks of the Gaussian fits, determined from the strongest line, $^{12}$CO(\mbox{3--2}), are listed in Table~\ref{t:sample}.

Because the lines of sight to the Galactic centre are crowded with interstellar CO line emission, the imaging was done twice for the CO lines, first with full range of visibilities and then excluding short baselines corresponding to spatial frequencies lower than \mbox{3--5}\,\arcsec, in order to exploit the filtering capabilities of the interferometer and limit the line-of-sight contamination from interstellar CO line emission. The latter approach was tested on a number of objects with strong circumstellar lines and no interstellar contamination and showed significant amounts of lost circumstellar flux in all cases. Hence, except for two objects, we refrained from using short-baseline removal as a solution to this problem. It should be noted that interstellar line contamination is, in general, less significant in the case of $^{13}$CO, where, in a relative sense, the circumstellar emission is stronger compared to the interstellar emission. Brief discussions on individual objects, affected by interstellar CO line emission, are given in Appendix~\ref{a:source_disc}, as well as of the three objects for which no circumstellar CO line emission was detected.

%
\subsection{APEX data}

For a sub-sample, the APEX 12\,m telescope \citep{gustetal06}  was used to observe the $^{12}$CO $J$\,=\,\mbox{2--1} (23), \mbox{3--2} (16), and \mbox{4--3} (13)  lines at 230, 345, and 461\,GHz, respectively, and the $^{13}$CO $J$\,=\,\mbox{2--1} (23) and \mbox{3--2} (16) lines at 220 and 330\,GHz, respectively (the number in parentheses gives the number of objects observed). The nFLASH230 heterodyne instrument (nFLASH built by Max-Planck-Institut f{\"u}r Radioastronomie) was used together with a Fast Fourier Transform Spectrometer (FFTS) covering 8\,GHz per sideband and polarisation for the \mbox{2--1} line observations in May and October 2023, August 2024, and April and August 2025 under programmes O-0111-9304 and O-0113-9300. The SEPIA345 heterodyne instrument \citep{meleetal22} was used together with an  FFTS covering 8\,GHz per sideband and polarisation for the \mbox{3--2} line observations in April and August 2024 and April 2025 under programme O-0113-9300. Finally, the nFLASH460 heterodyne instrument was used together with an  FFTS covering 4\,GHz per sideband and polarisation for the $^{12}$CO(\mbox{4--3}) line observations in August 2024 and April and August  2025 under programme O-0113-9300. 

The observations were made in a dual-beamswitch mode where the source is alternately placed in the signal and the reference beam, using a beam throw of about 1\arcmin. Regular pointing checks were made on strong CO line emitters and continuum sources. Typically, the pointing was found to be consistent with the pointing model within $\approx$\,3\arcsec. 

The raw spectra are stored in $T_{\mathrm A}^{\star}$-scale, where the antenna temperature is corrected for atmospheric attenuation using a calibration unit. The uncertainty in the absolute intensity scale is estimated to be about $\pm$\,10\,\%. The data were reduced by removing a first-order polynomial baseline using the CLASS/GILDAS software package\footnote{\url{https://www.iram.fr/IRAMFR/GILDAS/doc/html/class-html/node38.html}} and further analysed using XS\footnote{XS is a package developed by P. Bergman to reduce and analyse single-dish spectra. It is publicly available from {\tt ftp://yggdrasil.oso.chalmers.se}}. The CO lines are expected to be essentially unpolarised, so the data for the two independent polarisations were averaged. Finally, the spectra were converted to flux density scale using the values 35, 36, and 46 Jy\,K$^{-1}$ at 230, 345, and 461\,GHz\footnote{\url{http://www.apex-telescope.org/telescope/efficiency/index.php}}, respectively, and the velocity resolution was reduced to 2\,km\,s$^{-1}$.

%
\subsection{Photometry and light curves}
\label{s:photometry}

Using the Vizier Photometry tool, we concluded that the infrared was the best wavelength range for characterising the SEDs.  We used the flux densities at 4.49 and 7.87\,$\mu$m observed using the Spitzer Space Telescope (Spitzer), at 8.28, 14.65, and 21.34\,$\mu$m using the Midcourse Space Experiment (MSX), and at 22.1~$\mu$m using the Wide-field Infrared Survey Explorer (WISE), to characterize the sample.

The reddening $E$($J$\,$-$\,$K_{\rm s}$) towards each source was calculated using the results of \citet{gonzetal18} and converted into extinction in the $K_{\rm s}$ band, $A_{K_{\rm s}}$, and extrapolated up to 8\,$\mu$m using the relationships provided by \citet{nishetal09}. In the range 8 to 30\,$\mu$m we use the extinction law provided by \citet{xueetal16}. The extinction corrections were applied to the flux densities used in Sect.~\ref{s:dust_cse_char}.

We estimated variability periods using data from the VISTA Variables in the V\'{i}a L\'{a}ctea (VVV) ESO Public Survey \citep{minnetal10} data release 5, and data from the combined WISE \citep{wrigetal10} and the Near-Earth Object WISE \citep[NEOWISE;][]{mainetal11,mainetal14} survey including data up to September 2023. The data covers a time span of about 5000$^{\rm d}$. Using the ALMA-based coordinates nearby counterparts within 6\arcsec\ were retrieved. For the counterpart, the available data in the $Z$ (0.88\,$\mu$m), $Y$ (1.02\,$\mu$m), $J$ (1.25\,$\mu$m), $H$ (1.64\,$\mu$m), and $K_{\rm s}$ (2.15\,$\mu$m) filters were retrieved. In $ZYJH$ there are typically four or fewer observations, while dozens of data points are available in $K_{\rm s}$. The $K_{\rm s}$-band light curves were analysed using the codes and methodology outlined in \citet{groeetal20} assuming a single period. For the WISE data a very similar approach was adopted to analyse the light curves in the W1 (3.4\,$\mu$m) and W2 (4.6\,$\mu$m) bands, as outlined in \citet{groe22}, also assuming a single period. In a few cases saturation in the $K_{\rm s}$ band is an issue, but periods could be estimated using the W1 and W2 band data. The results, including the adopted periods, are given in Table~\ref{t:variability_results}.

%
%
%
\section{Object categories}
\label{s:categories}

Before presenting the results, we divide our objects into categories based on three observational characteristics: the SED, the CO line emission, and the near-infrared variability. Remarkably, as will be shown below, for each of these characteristics our objects clearly divide, with only a few exceptions, into two types: attached (`standard') versus detached dust-CSEs, standard versus complex gas-CSEs, and periodic versus non-periodic light curves. Based on these three bi-modal distributions, we divide our objects into four categories in Sect.~\ref{s:abcd}. This division into categories of objects with different stellar and/or circumstellar characteristics, will guide the way the observational results are presented in this paper, and how they are further analysed in upcoming papers in preparation.

%
\subsection{Dust-CSE characteristics}
\label{s:dust_cse_char}

The dust-CSE characterisation is based on the SED morphology. A visual inspection shows that the sample clearly separates into two types of SEDs. Although all stars show strong emission in the mid-infrared, one type containing most objects also shows strong emission at $\approx$\,5\,$\mu$m, while the remaining objects are much dimmer at these shorter wavelengths. SED model-fitting tests (based on spherical dust-CSE with $r^{-2}$ dust density distribution) show that the more reddened objects have relatively large inner radii of their dust-CSEs ($>$\,100\,au), these we consider to have a detached envelope (DdE), while the rest have a `standard' dust-CSE (SdE; inner radius of about 10\,au).

\begin{figure}[h!]
   \centering
     \includegraphics[width=7cm]{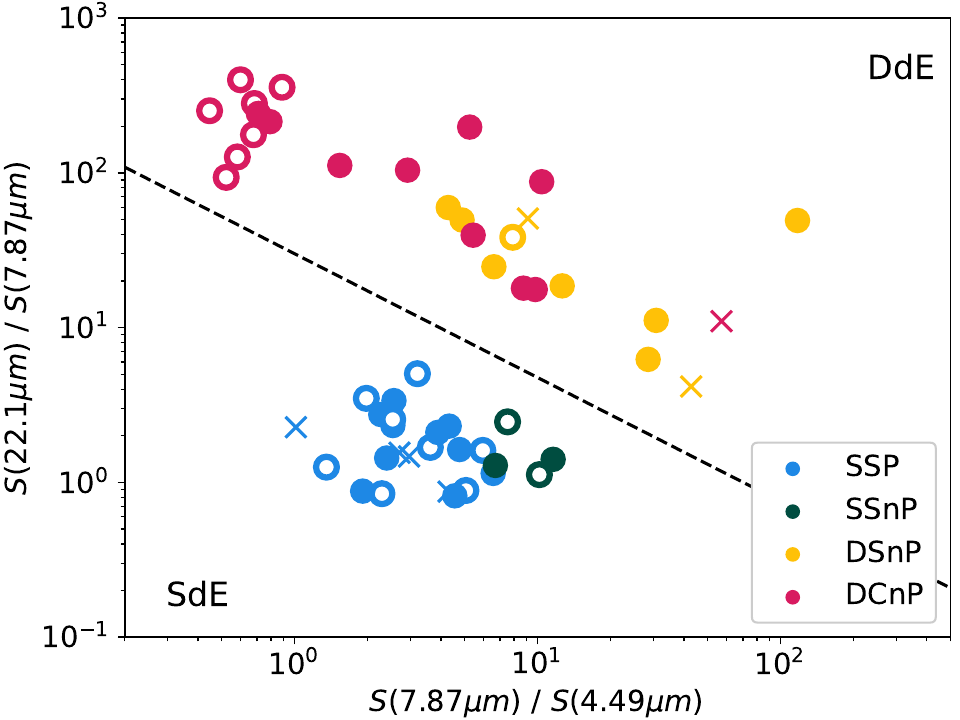}   \includegraphics[width=7cm]{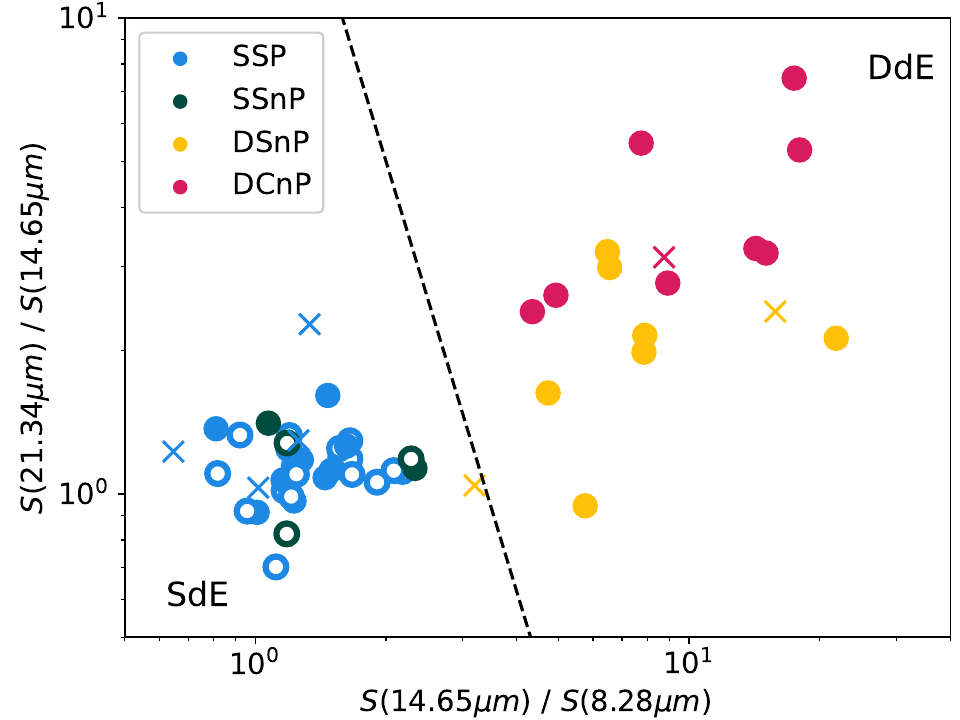}
      \caption{Sample objects in two colour-colour diagrams based on Spitzer + WISE (top) and MSX (bottom) data. Details of the lines separating the objects into types, SdE (blue and green) and DdE (yellow and cyan), are given in the text. Objects with a solid circle appear in both plots. For objects marked with crosses, some of the flux densities are interpolated or extrapolated. The categories introduced in Sect.~\ref{s:abcd} are used here.}
         \label{f:colour_colour}
   \end{figure}

In fact, full SED modelling is not required to separate the sources into SdE and DdE types. To quantify the difference between the two SED types, we used the extinction-corrected flux densities at 4.49 and 7.87\,$\mu$m (Spitzer), at 8.28, 14.65, and 21.34\,$\mu$m (MSX), and at 22.1\,$\mu$m (WISE). Based on these data, we constructed two colour-colour diagrams: $S_{7.87}/S_{4.49}$ vs $S_{22.1}/S_{7.87}$ and  $S_{14.65}/S_{8.28}$ vs $S_{21.34}/S_{14.65}$. Of the 74 sources detected in CO line emission, 67 have the required observations to be placed in at least one of these diagrams, marked as solid circles in Fig.~\ref{f:colour_colour}. As shown in Fig.~\ref{f:colour_colour}, the two SED types are easily distinguishable based on their infrared colours. The dashed lines in the diagrams are indicative of the borders between the two types. These borders are mathematically described by the expressions $S_{22.1}/S_{7.87}$\,=\,30$\times$($S_{7.87}/S_{4.49})^{-0.8}$ and $S_{21.34}/S_{14.65}$\,=\,40\,$\times$\,($S_{14.65}/S_{8.28})^{-3}$. The remaining seven sources, marked as crosses in Fig.~\ref{f:colour_colour}, lack at least two of the required values. To add these sources to the diagrams, we estimated the flux densities at the wavelengths with missing data by interpolating the SED. In the case of OH000.319-0.041, no photometric points are available beyond 14.5\,$\mu$m and we assumed the SED to be flat up to 22.1\,$\mu$m. Although this object lies between the two types in the $S_{14.65}/S_{8.28}$ vs $S_{21.34}/S_{14.65}$ diagram, we characterize it as DdE because its $S_{7.87}/S_{4.49}$ colour is quite red compared to those of the SdE objects.

In total, there are 47 and 27 objects of type SdE and DdE, respectively. The SdE objects are confined to relatively small regions in both diagrams, suggesting a homogeneous set of dust-CSEs. The DdE objects, on the other hand, are more spread out indicating a more heterogenous set of dust-CSE properties.

   \begin{figure*}[t]
   \centering
    \includegraphics[width=16cm]{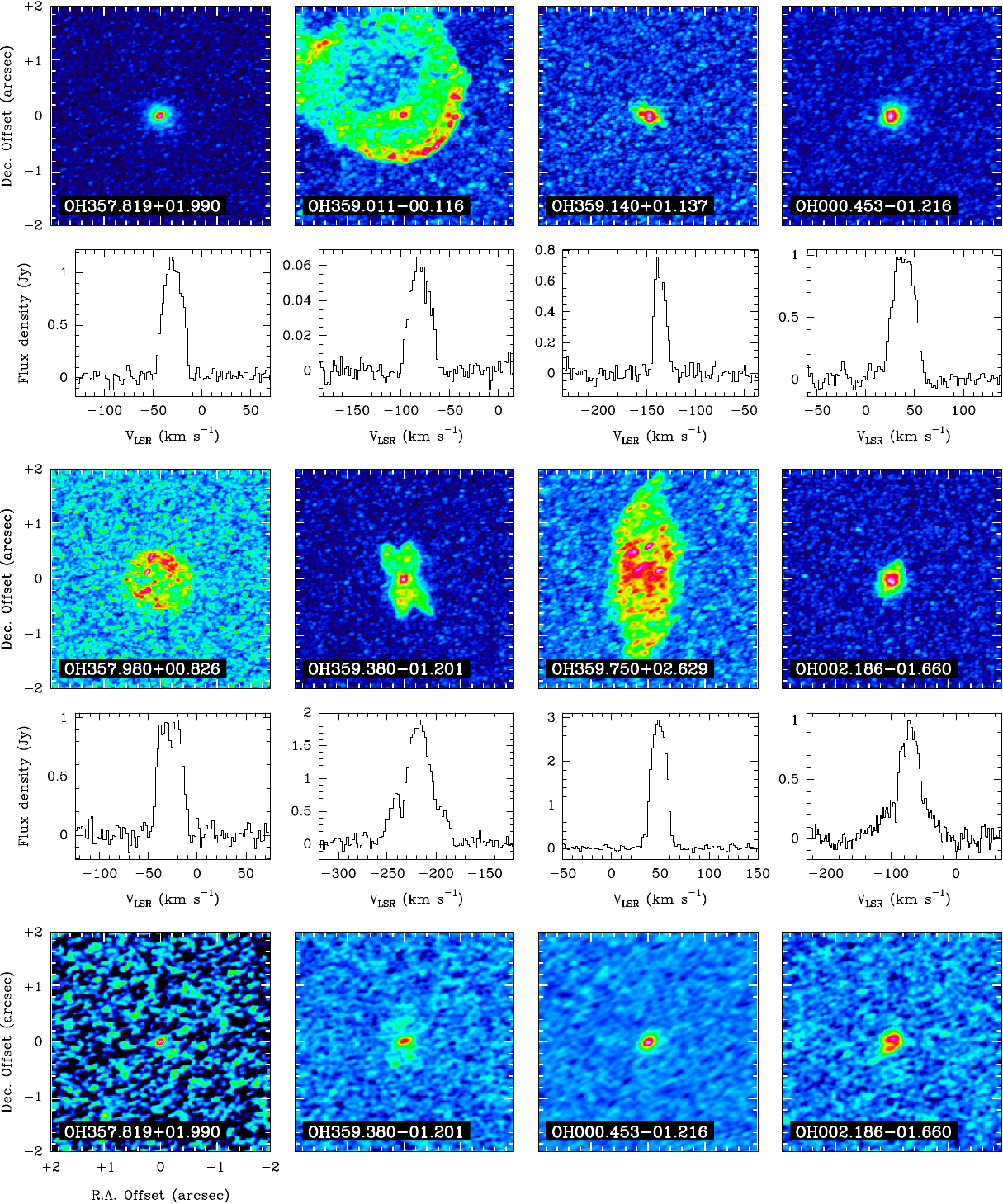} 
     \caption{ALMA data. {\it First row:} Maximum-intensity $^{12}$CO(\mbox{3--2}) images of objects with centrally peaked brightness distributions and with only weak extended emission (OH359.011$-$0.116 is suspected to be affected by CSM/ISM interaction; see text for details). {\it Second row:} Line profiles of $^{12}$CO(\mbox{3--2}) for the same objects (aperture 2\arcsec, except OH359.011$-$0.116 for which 0\farcs15 is used). {\it Third row:}  Maximum-intensity $^{12}$CO(\mbox{3--2})  images of objects with extended, complex structure. {\it Fourth row:} Line profiles of $^{12}$CO(\mbox{3--2}) for the same objects (3\arcsec\ aperture). {\it Fifth row:} Images of continuum at 338\,GHz.}
         \label{f:obs_examples}
   \end{figure*}
%
%
%
\subsection{Gas-CSE characteristics}
\label{s:gas_cse_char}

The gas-CSE characterisation is based on the large-scale morphology of the brightness distribution and shape of the $^{12}$CO lines. We extracted $^{12}$CO line profiles as presented in Sects~\ref{s:line_prof_abc} and \ref{s:bright_prof_d} and inspected them by eye. The line shapes can be divided into the soft parabola type (66 objects) or  a central soft-parabola component complemented with prominent extended higher-velocity wings type (eight objects). We used maximum intensity $^{12}$CO images (moment 8 in CASA), over the circumstellar velocity ranges, to characterize their morphologies. The majority of the objects have $^{12}$CO line brightness distributions that are sharply peaked at the centre, with only weak extended emission without any discernible signs of structure. There are 59 objects of this type. There are 15 objects that show extended emission with complex structures. In most of these cases there  is no central peak, and the extended structures have an axial symmetry. Examples of brightness distributions and line profiles are shown in Fig.~\ref{f:obs_examples}. 

When combining these results, we find 57 objects with brightness distributions that are sharply peaked at the centre and have soft-parabola line profiles. These objects will be analysed using a standard gas-CSE (SgE; spherical CSE that is formed by a constant MLR, and that expands with a constant velocity). Among these, there are both SdE and DdE objects. The remaining 17 objects have complex brightness distributions, or, in the two cases with sharply centrally peaked brightness distributions, the line profiles have prominent extended line wings. These objects will be analysed using a more complex gas-CSE (CgE). These are all of the DdE type.  There may be objects with higher-velocity wings that are too weak to be detected or the orientation is such that the wings are not seen if the outflow is bipolar, i.e. there may be border line objects between SgE and CgE.

We note here that about 20\,\% of the objects show surrounding structure, in both $^{12}$CO and $^{13}$CO, that suggests an interpretation in the form of CO line emission coming from a region where expanding circumstellar gas interacts with the surrounding interstellar medium. In at least seven cases, there is a bow-shock structure with the star at the expected position if this is due to emission from interacting gas caused by stellar motion through the ISM. A clear example is the OH359.011$-$0.116 $^{12}$CO(3$-$2) image in Fig.~\ref{f:obs_examples}. This phenomenon will be analysed in a separate paper (Maercker et al., in prep.).

%
%
   \begin{figure}[t]
   \centering
   \includegraphics[width=7cm]{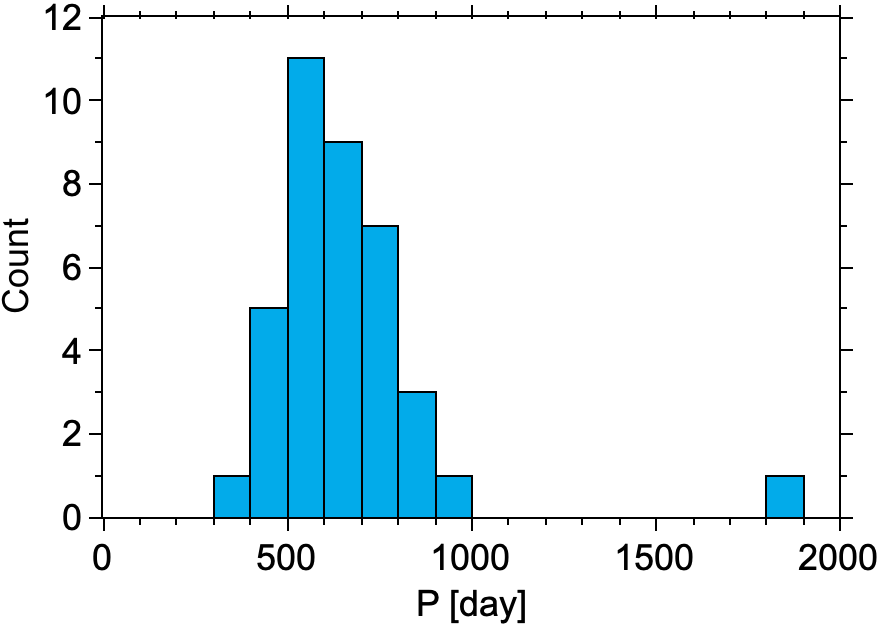}  
     \caption{Distribution of periods for the objects with large amplitude, regular variability.}
         \label{f:p_a}
   \end{figure}

%
\subsection{Variability characteristics}
\label{s:variability}

The variability characterisation is based on the near-infrared light curve analysis presented in Sect.~\ref{s:photometry}. We estimated pulsation periods and amplitudes for the $K_{\rm s}$ and/or WISE bands, see Table~\ref{t:variability_results}.  Each light curve is also visually inspected. We regard an object as a large-amplitude, regular pulsator (P) if the period is well-determined and the amplitude is $>$\,0.5\,mag, with an uncertainty lower than 20\,\% of its value, in at least one band. There are 40 objects of this type. For them we also give an adopted period, estimated as the average of the periods in the different bands weighted with the uncertainties, Fig.~\ref{f:p_a}. The median period of this sub-sample is 613$^{\rm d}$.  There also are 34 objects where there is enough data to conclude that the light curves are clearly irregular, or the estimated amplitudes are low ($\la$\,0.1\,mag) (nP). This shows a distinct division into objects of type P and nP. Only in one case there is not enough data to draw any conclusions on the variability, and in two cases the P identification is only tentative. Comparing these results with those of the SED and CO line types, we find that the 40 P-type objects are all of SdE and SgE type, while the 34 nP type objects are a mixture of all types. The one object for which there is not enough data to determine the variability is of CgE type.

\begin{table}[t]
\caption{Division of objects into categories.}           
\label{t:categories}    
\centering                
\begin{tabular}{l l c c c}      
\hline\hline 
\\[-2ex]      
Category  & Characteristics\,$^a$                      & \#\,$^b$                      & C$^{18}$O\,$^c$       &  338\,GHz\,$^d$      \\        
\hline 
\\[-2ex]                    
   SSP      & SdE, SgE, P                                     &  40                              & \phantom{1}2 (0.05)   & 11 (0.28)     \\    
   SSnP    & SdE, SgE, nP                                   &   \phantom{0}7            & \phantom{1}0 (0.00)   & \phantom{1}4 (0.57)        \\
   DSnP    & DdE, SgE, nP                                   &  10                              &\phantom{1}3 (0.30)    & \phantom{1}8 (0.80)      \\
   DCnP    & DdE, CgE, nP                                   &   17                             & 11 (0.65)                    & 10 (0.59)                         \\
\hline                                 
\end{tabular}
\tablefoot{$^{(a)}$\,SdE = standard dust-CSE, DdE = detached dust-CSE, P = regular, large-amplitude variability, nP = low-amplitude or no regular variability, SgE = standard gas-CSE, CgE = complex gas-CSE. $^{(c)}$\,Number of objects in each category. $^{(c)}$\,Number of objects detected in the C$^{18}$O(\mbox{2--1}) line, and the occurrence rate in parentheses. $^{(d)}$\,Number of objects detected in continuum at 338\,GHz, and the occurrence rate in parentheses.}
\end{table}

%
\subsection{Division into categories}
\label{s:abcd}

We use the three characteristics presented above to divide our objects into four categories of distinct stellar and/or circumstellar characteristics. There are 40 objects in category SdE/SgE/P (SSP), seven objects in category SdE/SgE/nP (SSnP), ten objects in category DdE/SgE/nP (DSnP), and 17 objects in category DdE/CgE/nP (DCnP).  The number of objects for which the category membership is uncertain are very few, as described above. The results are summarised in Table~\ref{t:categories}, and the category of each object is given in Table~\ref{t:sample}. The evolutionary phases along the AGB and beyond corresponding to these categories will be discussed in a forthcoming paper.

The colour-colour diagrams, Fig.~\ref{f:colour_colour}, add extra information of relevance for the introduced categories. In the $S_{7.87}/S_{4.49}$ vs $S_{22.1}/S_{7.87}$  diagram there is a separation between SSP and SSnP objects, although this is based on only four objects of the latter category. Further, in both colour-colour diagrams there is a tendency for DSnP and DCnP objects to separate into two different regions.

The detections of the C$^{18}$O(\mbox{2--1}) line (Sects~\ref{s:c18o_abc} and \ref{s:c18o_d}) and continuum at 338\,GHz (Sects~\ref{s:continuum_abc} and \ref{s:continuum_d}) add further information on to what extent objects in the various categories differ from each other. The number of objects detected with each of these characteristics is given in Table~\ref{t:categories}. The detection rate of the C$^{18}$O line is highest in category DCnP. The mm-wave continuum detection rate is lowest among the SSP objects.

%
%
\section{Observational results: ALMA, SgE objects}
\label{s:results_abc}

We start by presenting the ALMA $^{12}$CO and $^{13}$CO results for SgE objects (i.e. categories SSP, SSnP, and DSnP; 57 in total), since these form a very homogeneous dataset in terms of their large-scale CO line and ALMA continuum characteristics, and they will be analysed separately in a forthcoming paper.

%
\subsection{Azimuthally averaged CO line radial brightness profiles}
\label{s:aarp}

\begin{figure}[h!]
   \centering
   \includegraphics[width=6.5cm]{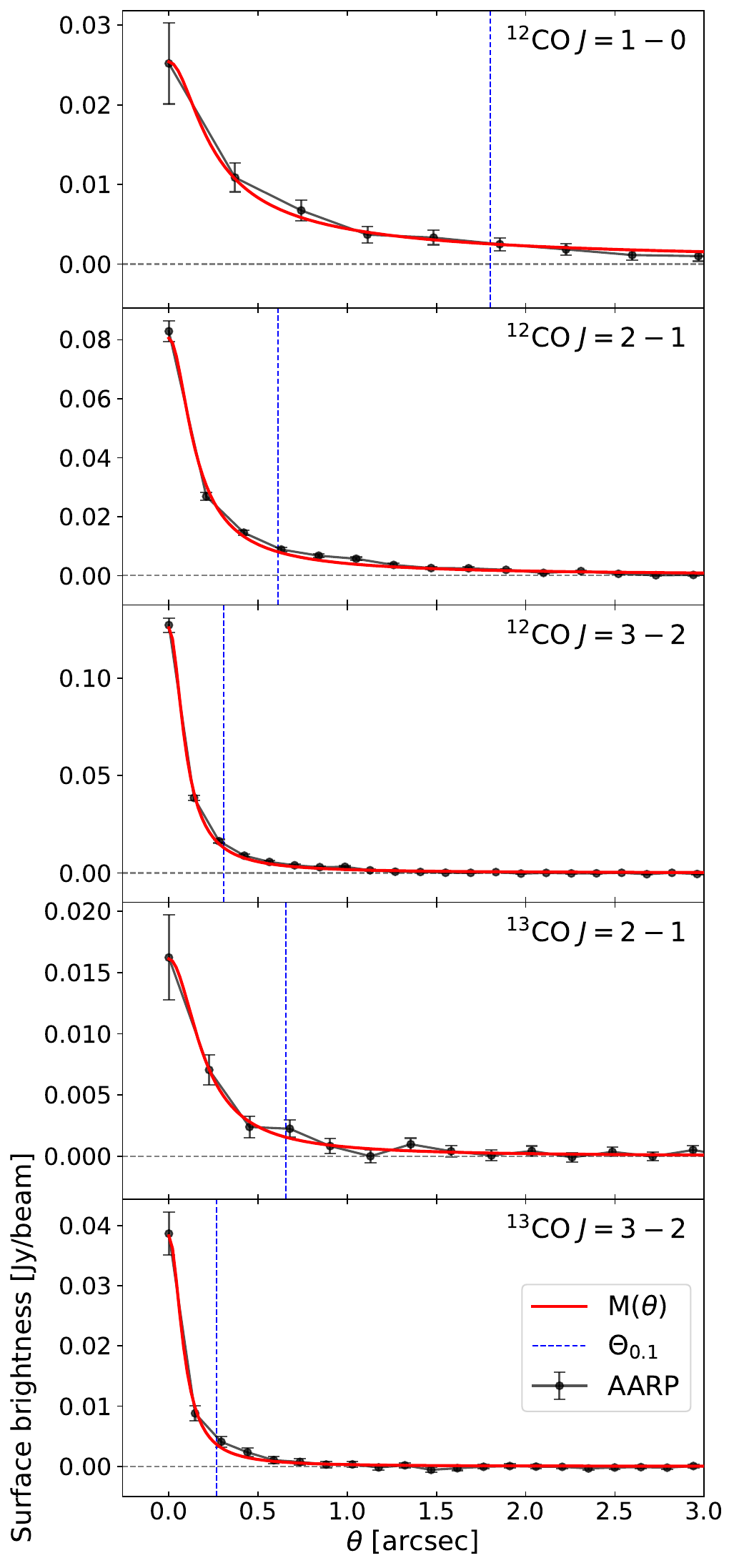}  
     \caption{Azimuthally averaged radial brightness profiles of OH357.819+1.990. The error bar of each data point is the $\pm$\,1$\sigma$ uncertainty. The Moffat fits are shown as red lines, and the $\Theta_{0.1}$:s are shown by the blue dotted lines (see text for details).}
         \label{f:aarps}
   \end{figure}

The SgE objects all have sharply, centrally peaked CO line brightness distributions, but also weak extended emission without obvious structure. In order to study them in more detail we derived azimuthally averaged radial brightness profiles (AARPs) by azimuthally averaging the data cubes for all objects and for all lines within the velocity range $\pm$20\,\% of the gas expansion velocity centred on the systemic velocity. This increases the S/N and provides more reliable information on the weak, extended emission. Before averaging, the data cubes are convolved with a circular Gaussian having a FWHM equal to the FWHM of the major axis of the synthesised beam. The resulting maps were divided into concentric circular annuli of increasing radii, centred on the estimated source position, with widths equal to that of the convolving beam. The resulting average brightness profile constitutes an AARP.  Examples of AARPs are given in Fig.~\ref{f:aarps}.

The resulting AARPs, for each individual line, all appear very similar in shape and they are clearly much more extended than the Gaussian synthesised beam. In fact, very good fits to the AARPs are obtained using a function introduced by \citet{moff69} to study extended focal stellar images: 
%
\begin{equation}
\label{e:moffat}
\Sigma(\theta) =  \frac{\Sigma(0)}{[1+(\theta/\alpha)^2]^\beta} 
\end{equation}
%
where $\alpha$ and $\beta$  are measures of the extent of the function ($\beta$\,=\,1 corresponds to a Lorentzian with a semi-width at half maximum (SWHM) equal to $\alpha$). The SWHM, $\Theta_{0.5}$, of this profile is given by
%
\begin{equation}
\label{e:R0.5}
\Theta_{0.5}  = \alpha \sqrt{2^{1/\beta} - 1}\,\, .
\end{equation}
%
The best-fit results, obtained through a $\chi^2$-analysis, are summarised in Table~\ref{t:aarp_sizes}, and examples are shown in Fig.~\ref{f:aarps}. The $\Theta_{0.5}$ values are, on average, close to the FWHMs of the synthesised beams, and not a good measure of the extent of the brightness distribution \citep[see also ][]{ramsetal20}.

The surface integral $\int_0^\infty 2\pi \Sigma (\theta) \theta {\rm d}\theta$ converges only when $\beta$\,$>$\,1, and only for reasonable angular extents if $\beta$\,$\ga$\,2 (taking into account the fact that the CO photodissociation radii for our sources correspond to $\approx$\,1\arcsec).  The average $\beta$:s are all below 2, and we cannot use these fits to reliably estimate the total flux densities. We instead chose the angular extent at which the fit to the AARP reaches 10\,\% of its peak value, $\Theta_{0.1}$, as a measure of the size of the brightness distribution (see Table~\ref{t:alma_line_continuum} for individual results and Table~\ref{t:aarp_sizes} for a summary). This will be compared with radiative transfer results in a forthcoming paper. It is clear that the dataset is very homogeneous (the scatter around the average is in all cases small), implying very similar large-scale circumstellar properties for all SgE objects. In fact, a significant contribution to the spread in the data is most likely a dependence of the brightness distribution on the MLR, the larger the MLR the larger the brightness distribution is expected to be, and we estimate that our sample spans at least one order of magnitude in MLR, based on the range of OH 1612 MHz flux densities. 

On average, for the $^{12}$CO lines, the extent of the 1$-$0 emission is $\approx$\,80\,\% larger than that of the 2$-$1 emission, which in turn is $\approx$\,60\,\% larger than the  3$-$2 emission. This is as expected since higher-$J$ transitions are more effectively excited closer to the star. For $^{13}$CO, the 2$-$1 emission extent is about 65\,\% larger than that of the  3$-$2 emission, and, on average, the $^{12}$CO emissions are about 30\,\% larger than the corresponding $^{13}$CO emissions.

\begin{table*}[h]
\caption{Results of the fits to the CO line AARPs  for SgE objects.}           
\label{t:aarp_sizes}    
\centering                
\begin{tabular}{l c c c c c}      
\hline \hline
\\[-2ex]  
                                            & $^{12}$CO(1$-$0)               &   $^{12}$CO(2$-$1)            &  $^{12}$CO(3$-$2)              &   $^{13}$CO(2$-$1)        &     $^{13}$CO(3$-$2)   \\
\hline                       
 \\[-2ex] 
 $\beta$                               & 1.38\,$\pm$\,0.78(35)        & 1.07\,$\pm$\,0.43 (44)        &  1.31\,$\pm$\,0.56 (45)        & 1.30\,$\pm$\,1.12 (34)   &    1.78\,$\pm$\,1.32 (37)\\  
 \\[-1.5ex]
 $\Theta_{0.5}$ [\arcsec]     & 0.37\,$\pm$\,0.13(35)        & 0.18\,$\pm$\,0.04 (44)        &  0.13\,$\pm$\,0.03 (45)        & 0.16\,$\pm$\,0.05 (34)   &    0.11\,$\pm$\,0.04 (37)\\ 
  \\[-1.5ex] 
$\Theta_{0.1}$ [\arcsec]      &  1.01\,$\pm$\,0.33 (40)        & 0.57\,$\pm$\,0.14 (45)        &  0.36\,$\pm$\,0.07 (45)        & 0.46\,$\pm$\,0.14 (40)   &    0.29\,$\pm$\,0.08 (43)\\   
\hline
\end{tabular}
\tablefoot{Results are given as averages and standard deviations. The number in parentheses gives the number of objects.}
\end{table*}
%

%
%
%
%
\subsection{Line profiles}
\label{s:line_prof_abc}

The CO line brightness distributions are sharply peaked for SgE objects, but they contain extended low-intensity emission as shown above. This, together with the limited S/N of the data and the fact that the S/N decreases for a compact source when using increasingly larger apertures \citep{tsuketal23}, makes it difficult to estimate the total intensities of the interferometer data. We therefore summarise the observational $^{12}$CO and $^{13}$CO line results in a way suitable for the radiative transfer analysis of SgE objects. To this end we convolved the data cubes so that they correspond to data observed with circular Gaussian beams of FWHMs of 0\farcs5, 1\arcsec, 2\arcsec, and 3\arcsec\ (henceforth we refer to these as the 0\farcs5, 1\arcsec, 2\arcsec, and 3\arcsec\ `aperture' data). In terms of spatial extents, half the FWHMs correspond to 3.1$\times$10$^{16}$, 6.0$\times$10$^{16}$, 1.3$\times$10$^{17}$, and 1.8$\times$10$^{17}$\,cm at the distance of the Galactic centre.

Using these convolved data cubes we extracted the line profiles from the pixels corresponding to the source positions given in Table~\ref{t:sample}. For the majority of the objects, and for most of the lines, this procedure works fine, resulting in sub-data-sets containing emission from increasingly larger angular scales. In some cases the S/N of the results obtained for the larger apertures are low (as explained above), or they are affected by interstellar CO line emission. In such cases we report results only up to the largest angular scale that gives reliable data. In a few cases, even the 0\farcs5 data are severely affected by contamination from interstellar CO line emission. In these cases we report the results obtained when convolving the data with Gaussians having FWHMs as small as about 0\farcs2.

For SgE objects, centre velocities ($\varv_{\rm c}$; in the local standard of rest frame), full widths at zero power  (FWZPs), $\Delta \varv$, and velocity-integrated flux densities, $I$, are determined through fitting the `Shell' profile 
%
\begin{equation}
\label{e:line_shape}
S(\varv) = S(\varv_{\rm c})\,\left[ 1 + 4H \left( \frac{\varv - \varv_{\rm c}}{\Delta \varv} \right)^2  \right] \,\, ,
\end{equation}
%
as defined in the CLASS/GILDAS software package, to the full dataset. The flux densities, $S(\varv_{\rm c})$, are given by $I/[\Delta \varv (1+H/3)]$. $H$, the horn-to-centre parameter, gives the line-shape information ($H$\,=\,--1 and 0 mean a parabolic and a rectangular line shape, respectively). The line flux densities and the corresponding velocity-integrated flux densities are given in Table~\ref{t:alma_line_continuum}. This table also includes the estimated systemic, $\varv_{\rm sys}$, and gas expansion velocity, $\varv_\infty$, of each object, obtained as the averages of the individual line estimates $\varv_{\rm c}$ and $\Delta \varv$/2, respectively, for the $^{12}$CO and $^{13}$CO 2$-$1 and 3$-$2 lines of the 0\farcs5, 1\arcsec, and 2\arcsec\ datasets.

The flux density uncertainties are affected by a combination of the S/N, the presence of ISM CO line emission, the size of the `aperture', and the absolute flux calibration uncertainty, and thus it is difficult to determine. The values given here are obtained as averages of the flux density errors estimated in the line-profile fitting and should be regarded as guidelines. For the 0\farcs5 data the 1$\sigma$ uncertainties are 6, 7, and 10\,mJy in bands 3, 6, and 7, respectively. The noise level increases with increasing aperture due to non-white noise in the interferometer data \citep{tsuketal23}. We estimate increases by factors of 1.5, 2.9, and 4.5 in B3, 1.9, 4.0, and 7.2 in B6, and 1.9, 4.1, and 6.3 in B7,  when increasing the aperture from 0\farcs5 to 1\arcsec, to 2\arcsec, and to 3\arcsec, respectively. The total increases in flux density from 0\farcs5 to 3\arcsec\ are, on average, 4.7, 3.5, and 2.2 in the $^{12}$CO 1$-$0, 2$-$1, and 3$-$2 lines, respectively, see Table~\ref{t:int_ratio_sizes}. Hence, except for B3, the S/N decreases with increasing aperture. In addition, there is possibly resolved-out flux as discussed in Sect.~\ref{s:alma_apex_co}. The uncertainties in the expansion and centre velocities depend on the S/N and the presence of interstellar CO line contamination, and are estimated to be, on average, about $\pm$1\,km\,s$^{-1}$.

%
\subsubsection{Line intensity ratios: Different angular scales}

The velocity-integrated line intensities obtained for the 0\farcs5, 1\arcsec, 2\arcsec, and 3\arcsec\ data can be used as a measure of the radial brightness distribution, i.e. they provide information on how much of the flux measured by the interferometer lies within the different angular scales for each line, and serve as a complement to the AARP:s.  We derived three ratios of consecutive angular scales, $I$(1\arcsec )/$I$(0\farcs5), $I$(2\arcsec )/$I$(1\arcsec ), and $I$(3\arcsec )/$I$(2\arcsec ) for the $^{12}$CO and $^{13}$CO lines. The results are given in Table~\ref{t:int_ratio_sizes}. It is estimated that the 3\arcsec\ aperture data contains most of the emission  recoverable by the interferometer for all lines ($\ga$\,80\,\%).  Also here, the scatter around the averages is small.

\begin{table}[t]
\caption{Ratios of velocity-integrated line intensities obtained with consecutive `apertures' for SgE objects.}           
\label{t:int_ratio_sizes}    
\centering                
\begin{tabular}{l c c c}      
\hline\hline
\\[-2ex]      
Line          & $I$(1\arcsec )/$I$(0\farcs5)    & $I$(2\arcsec )/$I$(1\arcsec )         & $I$(3\arcsec )/$I$(2\arcsec )   \\        
\hline
\\[-1.5ex] 
                    &                                                \multicolumn{3}{c}{$^{12}$CO} \\ 
\cline{2-4} 
\\[-2ex]                   
   1--0         & 2.16\,$\pm$\,0.38 (46)                  &  1.78\,$\pm$\,0.33 (43)                        & 1.25\,$\pm$\,0.20 (38)  \\    
   2--1        & 1.93\,$\pm$\,0.26  (46)                  &  1.54\,$\pm$\,0.24 (43)                        & 1.15\,$\pm$\,0.12 (38)\\
   3--2        & 1.63\,$\pm$\,0.18 (45)                   &  1.28\,$\pm$\,0.16 (42)                        & 1.04\,$\pm$\,0.10 (34)  \\
   \\[-1.5ex]
                   &                                                \multicolumn{3}{c}{$^{13}$CO}  \\
\cline{2-4}
\\[-2ex]  
    2--1        & 1.71\,$\pm$\,0.28 (51)                  &  1.41\,$\pm$\,0.22 (46)                       & 1.18\,$\pm$\,0.17 (34)\\
   3--2        & 1.47\,$\pm$\,0.19 (52)                   &  1.25\,$\pm$\,0.20 (38)                       & 1.07\,$\pm$\,0.17 (24)  \\      
\hline                                 
\end{tabular}
\tablefoot{Results are given as averages and standard deviations (including the line intensity uncertainties). The number in parentheses gives the number of objects.}
\end{table}

%
\subsubsection{Line intensity ratios: Different transitions and isotopologues}

\begin{figure}[h!]
   \centering
   \includegraphics[width=7cm]{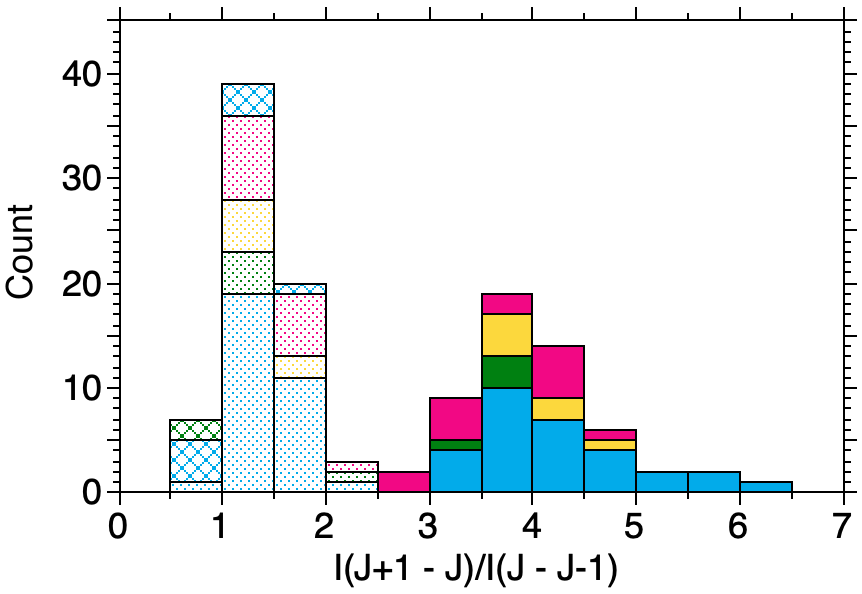}  
     \caption{Line intensity ratio distributions of $^{12}$CO for SSP (blue), SSnP (green), DSnP (yellow), and DCnP (cyan) objects plotted on top of each other. The line intensity ratios are (2$-$1)/(1$-$0) (solid), (3$-$2)/(2$-$1) (dotted), and (4$-$3)/(3$-$2) (crossed). }
     \label{f:line_ratios_abcd}    
   \end{figure}

We use the velocity-integrated line intensities to calculate line intensity ratios between consecutive transitions, $I$($J$+1\,$\rightarrow$\,$J$)/$I$($J$\,$\rightarrow$\,$J$-1), for SgE objects. This was done only for those sources that have reliable data up to a spatial scale of 2\arcsec. An average value for the three data points for each line ratio and each source was calculated. The results are given in Table~\ref{t:int_ratios_transitions} and Fig.~\ref{f:line_ratios_abcd} for the $^{12}$CO lines. We also include the $^{12}$CO \mbox{4--3/3--2} line intensity ratio from the APEX data. The dataset is characterized by only small scatter around the averages. The \mbox{2--1} line is significantly stronger than the \mbox{1--0} line, by about a factor of four, and the \mbox{3--2} line is somewhat stronger than the \mbox{2--1} line, by about a factor of 1.5, while the \mbox{3--2} and \mbox{4--3} lines have about equal strength. It should be noted that these ratios are expected to depend on the MLR (i.e. depend on optical depth), and somewhat higher values should apply for objects of MLRs lower than typical for our objects.

We calculated the $^{12}$CO/$^{13}$CO line intensity ratios for the 2$-$1 and 3$-$2 lines in the same way as above. They are surprisingly low, 1.5$-$8 with an average of 3.4\,$\pm$\,1.2 (the two lines combined). Even if they are likely to be somewhat affected by opacity effects \citep[see e.g.][]{sabeetal20}, the conclusion is that the circumstellar $^{12}$CO/$^{13}$CO abundance ratios are likely to be low. The $^{12}$CO/$^{13}$CO abundance ratios will be calculated, and discussed in relation to stellar $^{12}$C/$^{13}$C ratios, in a forthcoming paper.

\begin{table}
\caption{Ratios of velocity-integrated line intensities.}           
\label{t:int_ratios_transitions}    
\centering                
\begin{tabular}{l c c}      
\hline\hline
\\[-2ex]  
\multicolumn{3}{c}{SgE objects} \\
\cline{1-3}
\\[-1.5ex] 
Ratio                       &      $^{12}$CO                       &   $^{13}$CO \\ 
\hline
\\[-2ex]     
2--1/1--0                 &    4.05\,$\pm$\,0.70 (42) \\
3--2/2--1                 &    1.43\,$\pm$\,0.22 (42)       & 1.59\,$\pm$\,0.26 (37)  \\
4--3/3--2\,${^a}$     &    1.01$\pm$\,0.27 (10) \\
\\[-1.5ex]
Line                       &     $^{12}$CO/$^{13}$CO  \\
\cline{1--1} \cline{2-2} 
\\[-2ex]                            
2--1                      &    3.54\,$\pm$\,1.34 (42) \\
3--2                      &    3.14\,$\pm$\,1.12 (33) \\
\hline
\\[-1ex] 
\multicolumn{3}{c}{CgE objects} \\
\cline{1-3}
\\[-1.5ex] 
Ratio                       &      $^{12}$CO                       &   $^{13}$CO \\ 
\hline
\\[-2ex]     
2--1/1--0                 &    3.98\,$\pm$\,1.46 (15) \\
3--2/2--1                 &    1.41\,$\pm$\,0.37 (15)       & 1.56\,$\pm$\,0.29 (17)  \\
\\[-1.5ex]
Line                       &     $^{12}$CO/$^{13}$CO  \\
\cline{1--1} \cline{2-2} 
\\[-2ex]                           
2--1                      &    2.80\,$\pm$\,0.52 (15) \\
3--2                      &    2.46\,$\pm$\,0.67 (16) \\
\hline                                 
\end{tabular}
\tablefoot{Results are given as averages and standard deviations (including the line intensity uncertainties). The number in parentheses gives the number of objects. For SgE objects the 0\farcs5, 1\farcs0, and 2\farcs0 datasets are used. $^{(a)}$ This result is based on the APEX data.}
\end{table}

%
\subsubsection{Line shapes}

\begin{figure}[t]
   \centering
   \includegraphics[width=9cm]{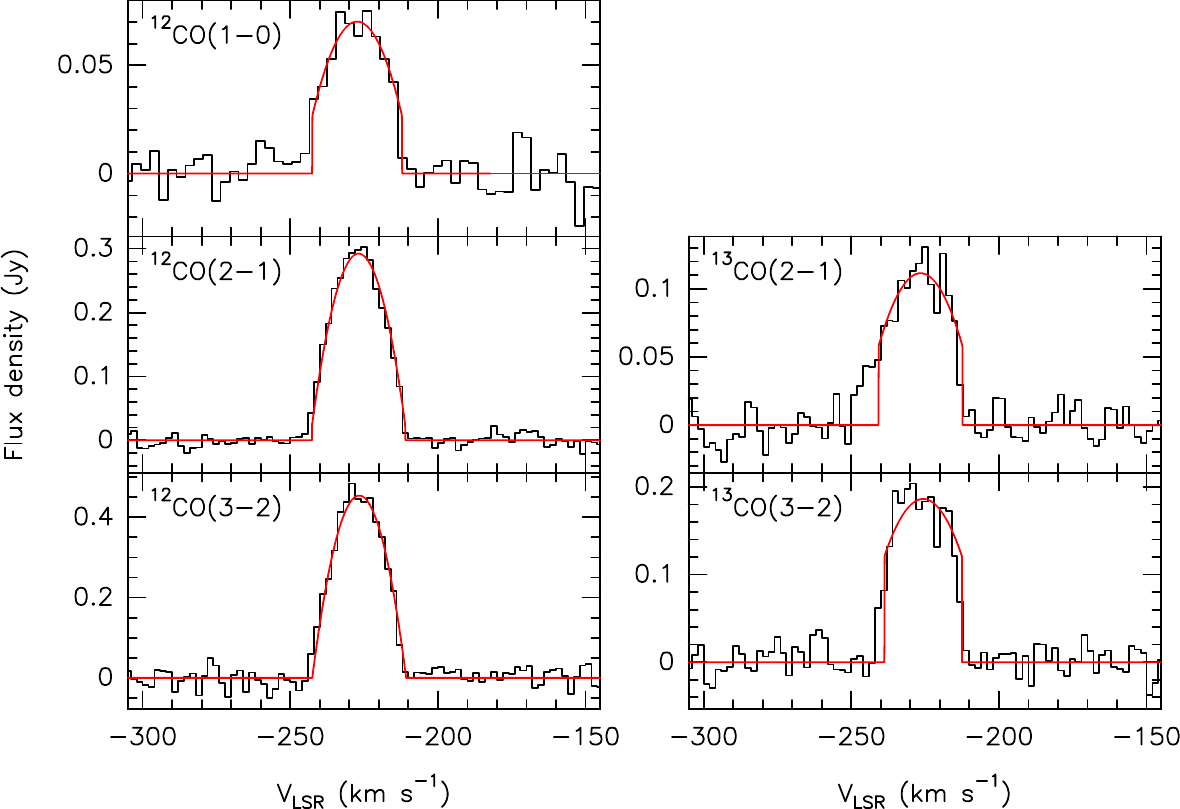}  
     \caption{Spectra of $^{12}$CO 1$-$0, 2$-$1, and 3$-$2, and $^{13}$CO 2$-$1 and 3$-$2 towards OH358.052+1.304 (black histogram) and the best shell-line-profile fits (Eq.~\ref{e:line_shape}, red solid line) with line-shape parameters $H$\,=\,$-$0.62, $-$0.92, $-$1.0, $-$0.48, and $-$0.35, respectively. The 1\arcsec\ data were used. This is an average object in our sample in terms of CO line intensity strengths.}
         \label{f:shell_fits}
   \end{figure}

Line shape information is obtained from the line fits. The $H$-values are sensitive to the S/N of the data, and to any contamination by CO line emission from the ISM and/or CSM/ISM interaction. We therefore refrained from tabulating results for each line and object. Table~\ref{t:h-values} summarises the statistical results for each line using the 0\farcs5, 1\arcsec, and 2\arcsec\ datasets (hence ignoring that the line shape is expected to depend somewhat on the aperture size if the emission is, at least, partly resolved), and an example is shown in Fig.~\ref{f:shell_fits}. Even though the scatter is significant, the trend is clear and follows what is expected considering line optical depths and extents of emission regions. The higher the energy of the transition, the higher is the optical depth and the smaller is the emission extent. Thus, it is expected that the $^{12}$CO lines stretch from close to rectangular ($H$\,=\,$-$0.22, the 1$-$0 line) to close to parabolic ($H$\,=\,$-$0.73, the 3$-$2 line), and the $^{13}$CO lines are more flat-topped than the corresponding $^{12}$CO lines.

\begin{table}
\caption{Line-shape data for SgE objects.}           
\label{t:h-values}    
\centering                
\begin{tabular}{l c}      
\hline\hline 
\\[-2ex]
Line                                & $H$ \\
\hline
\\[-2ex]
$^{12}$CO(1$-$0)          &    $-$0.22\,$\pm$\,0.38 (42)   \\
$^{12}$CO(2$-$1)          &   $-$0.66\,$\pm$\,0.28 (44) \\                 
$^{12}$CO(3$-$2)          &   $-$0.73\,$\pm$\,0.26 (45) \\
  \\[-1.5ex]
$^{13}$CO(2$-$1)          &   $-$0.17\,$\pm$\,0.49 (42)\\
$^{13}$CO(3$-$2)          &   $-$0.28\,$\pm$\,0.45 (39)\\
\hline
\end{tabular}
\tablefoot{The results are given as averages and standard deviations using the 0\farcs5, 1\farcs0, and 2\farcs0 datasets. The number in parentheses gives the number of objects.}
\end{table}

%
%
\subsection{Kinematics}
\label{s:kinematics_abc}

\begin{figure}[t]
   \centering
   \includegraphics[width=7cm]{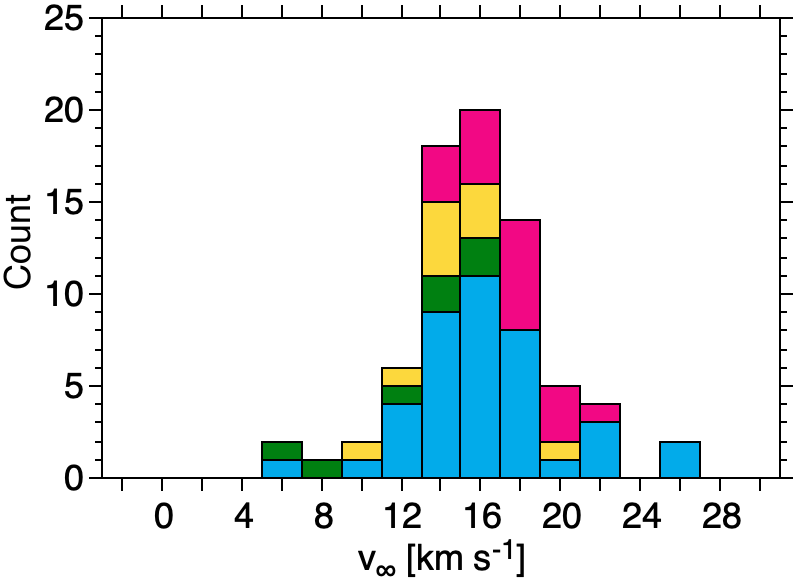}  
     \caption{Gas expansion velocities of the AGB CSEs of SSP (blue), SSnP (green), DSnP (yellow), and DCnP (cyan) objects plotted on top of each other.}
     \label{f:kinematics_abcd}
   \end{figure}

The gas expansion velocities from our line fits form a relatively narrow distribution, from 5 to 24\,km\,s$^{-1}$, Fig.~\ref{f:kinematics_abcd}. The average and standard deviation are 15\,$\pm$\,4\,km\,s$^{-1}$. This is a relatively high gas expansion velocity when compared to a median of 7\,km\,s$^{-1}$ for a solar neighbourhood sample of O-rich AGB stars \citep{ramsetal09}, but the range and average agree well with those of a larger sample of O-rich objects \citep{walletal25}. Further, it matches well the peaks of the velocity distributions of OH/IR stars in the Galactic plane, 12\,km\,s$^{-1}$, the Galactic bulge, 14\,km\,s$^{-1}$, and the Galactic centre, 19\,km\,s$^{-1}$, \citep{sjouetal98}, i.e. objects of similar MLRs as our objects. When compared to the expansion velocities estimated from the separation of the two OH 1612\,MHz maser peaks, the results agree well.

%
%
\subsection{C$^{18}$O}
\label{s:c18o_abc}

The C$^{18}$O(\mbox{2--1}) lines are substantially weaker than the corresponding $^{12}$CO and $^{13}$CO lines, as expected from a high $^{16}$O/$^{18}$O ratio (the solar value is $\approx$\,500, and it is not expected to change significantly during evolution for stars with a mass below $\approx$\,4\,$M_\odot$ \citep{karaluga16}. Among SgE objects, we detect only five objects (9\,\% detection rate), and the S/N of the detections are low. In Table~\ref{t:alma_line_continuum}, we report only the results from the 2\arcsec\ data, where the extracted line profiles were fitted with parabolic line shapes using the systemic and expansion velocities determined from the $^{12}$CO and $^{13}$CO lines. The use of the 2\arcsec\ data is a compromise between S/N and recovering as much of the line flux as possible.

%
\subsection{Continuum}
\label{s:continuum_abc}

We determined the positions and flux densities of the continuum by fitting  two-dimensional Gaussians to the data in the image plane. When detected, the continuum peak coincides with the CO line peak in all cases within $\approx$\,0\farcs1. The continuum brightness distributions are compact for SgE objects, with deconvolved Gaussian sizes smaller than the synthesised beam. Examples of continuum images are shown in Fig.~\ref{f:obs_examples}. The flux densities are summarised in Table~\ref{t:alma_line_continuum}. In total, two (4\,\% occurrence rate), ten (18\,\%), and 23 (40\,\%) SgE objects were detected in 107, 233, and 338\,GHz continuum, respectively. There is a marked difference between SSP, SSnP, and DSnP objects in terms of occurrence rate of the 338\,GHz continuum (28\,\%, 57\,\%, and 80\,\%, respectively), even if the number of objects are relatively low for the latter two categories. When detections are made in more than one band, the flux densities result in an average spectral index $\alpha$ ($S_\nu$\,$\propto$\,$\nu^\alpha$) of 3.5\,$\pm$\,0.6  (ten objects). The continuum flux at 338\,GHz of a 2500\,K blackbody of luminosity 5500\,$L_\odot$ (typical for our sources) at the distance of the Galactic centre is about 0.04\,mJy. This is well below our detection limit, and it, combined with the spectral index estimates, implies that any detected continuum emission from AGB stars at that distance is due to circumstellar dust emission. 

It is likely that there is continuum flux not captured by the fitted Gaussians nor `seen' by the interferometer. Extended low-surface brightness emission from colder dust at large radii may contribute (maybe even significantly) to the flux measured by ALMA. The MRS of our 338\,GHz observations is about 1\farcs9, corresponding to about 16000\,au at the distance of the Galactic centre, a distance covered by a 15\,km\,s$^{-1}$ outflow in about 2600\,yr. The high-MLR phase may have lasted longer than this, and consequently the dust-CSE may be larger than this. This will be further analysed in a forthcoming paper.
%
%
%
%
\section{Observational results: ALMA, CgE objects}
\label{s:results_d}

The CgE objects (category DCnP; 17 in total) form a much more heterogeneous dataset in terms of their CO line and ALMA continuum characteristics compared to SgE objects. The data will be presented here in a relatively concise form. They will be analysed separately in a forthcoming paper.

%
\subsection{Brightness distributions and line profiles}
\label{s:bright_prof_d}

We used the maximum-intensity images of the  $^{12}$CO line brightness distributions and the line shapes in order to summarise the characteristics of also these objects. The flux densities and kinematical information are retrieved from line profiles extracted using 3\arcsec\ apertures, or, in five cases, smaller apertures to minimize contamination from interstellar CO line emission. The line intensities are measured by fitting Gaussians to the obtained line profiles (in this way the objects are treated in the same way, irrespective of having line wings or not), the results are given in Table~\ref{t:alma_line_continuum}. The kinematical information (obtained through visual inspection), in terms of systemic velocity, expansion velocity of the soft-parabola component, and the FWZPs for the objects with line wings, is reported in Table~\ref{t:alma_line_continuum}. In summary, five objects have sharply centrally peaked brightness distributions with only weak extended emission, two objects have brightness distributions that appear as essentially circular rings ($\approx$\,1\arcsec\ in size) with no central peak, and the remaining ten objects have extended (0\farcs5 -- 4\arcsec\ in size) clearly axi-symmetric structures, see Fig.~\ref{f:obs_examples}. In most cases there is no central peak. Eight objects have CO line profiles with extended line wings.

We estimated line intensity ratios between consecutive transitions and between isotopologues also for CgE objects, see Table~\ref{t:int_ratios_transitions}. The results are very similar to those of the SgE objects (for CgE objects, the line intensities are dominated by the soft-parabola component).

%
\subsection{Kinematics}
\label{s:kinematics_d}

The gas expansion velocities of the central component of the line profiles of CgE objects are more uncertain, but their distribution appears similar to those of SgE objects, Fig.~\ref{f:kinematics_abcd}. When compared with the expansion velocities estimated from the separation of the OH 1612\,MHz maser peaks the results agree well. For CgE objects with line wings, the FWZPs range from about 50 to 150 \,km\,s$^{-1}$. The S/N of our spectra are in general limited, so the reported values are likely to be lower limits to the true FWZPs.

%
\subsection{C$^{18}$O}
\label{s:c18o_d}

The occurrence rate of the C$^{18}$O(\mbox{2--1}) line is much higher among CgE objects (65\,\%) compared to SgE objects (9\,\%).  Also for these objects the S/N of the detections are limited, and we report only the results from the 2\arcsec\ data, where the extracted line profiles were fitted using Gaussians, Table~\ref{t:alma_line_continuum}.

%
\subsection{Continuum}
\label{s:continuum_d}

The continuum brightness distributions of CgE objects often have a diffuse appearance covering about 0\farcs5. We determined the positions and flux densities by fitting  two-dimensional Gaussians to the data in the image plane. When detected, the continuum peak lies at the centre of the CO line emission. The flux densities are given in Table~\ref{t:alma_line_continuum}. In total one, eight, and ten CgE objects were detected in 107, 233, and 338\,GHz continuum, respectively. The occurrence rate of 338\,GHz continuum is higher than for SgE objects, 60\,\% versus 40\,\%, but not substantially higher. As for SgE objects, the relative occurrence rates in the three bands are fully explained by a steep spectral index (3.3\,$\pm$\,0.3, seven objects). Only one object is a clear exception: OH002.640$-$0.191 has a strong (in a relative sense), essentially flat continuum. 

%
%
%
\section{Observational results: APEX, all categories}
\label{s:alma_apex_co}

The objects observed with APEX were selected based on their systemic velocities (in an absolute sense, the highest ones) and location with respect to the Galactic plane (more than 1$^\circ$ above or below) to minimize the effect of contamination with interstellar CO line emission. The primary reason for the APEX observations was to provide estimates of flux not recovered by the interferometer for the $^{12}$CO and $^{13}$CO 2$-$1 and 3$-$2 lines. In total, 22 objects were detected in at least one CO line, see Table~\ref{t:apex_line_data}. The $^{12}$CO(4$-$3) line, not covered by our ALMA observations, was detected in 13 objects. 

To compare APEX and ALMA fluxes turned out to be somewhat complicated. The reason is that for about 20\,\% of the objects in the full sample there  is extended CO line emission that is most likely coming from an interaction region between the expanding circumstellar gas and the surrounding interstellar medium. This emission is often of relatively low surface brightness, but becomes significant when integrated over the full APEX beam, in some cases even dominating over the circumstellar emission. We therefore removed the sources, where such a contamination can be suspected, from the comparison. The resulting ratios between the ALMA (3\arcsec\ data) and  APEX velocity-integrated CO line intensities for each line are given in Table~\ref{t:alma_apex}. The data from all categories are treated equally.

We do not expect substantial loss of flux in these lines, since the extents of the emissions are clearly smaller than the MRSs (2$\Theta_{0.1}$/MRS\,$\approx$\,0.3$-$0.5). Taken at face value the results indicate $\approx$ 10 and 20\,\% loss of flux in B6 and B7, respectively. However, the APEX lines may include contributions from extended low-intensity emission, and the line intensities should be seen as upper limits. Taking everything into consideration, the results indicate that the typical flux loss in the ALMA data is at the $\la$\,20\,\% level.

\begin{table}
\caption{Ratios of velocity-integrated line intensities between ALMA and APEX data.}           
\label{t:alma_apex}    
\centering                
\begin{tabular}{l c c}      
\hline\hline 
\\[-2ex] 
Line                   &      \multicolumn{2}{c}{ALMA(3\arcsec)/APEX} \\
\cline{2-3} 
\\[-1.5ex] 
                          &      $^{12}$CO                       &   $^{13}$CO \\ 
\hline 
\\[-2ex]    
2--1              &    0.87\,$\pm$\,0.22 (18)       & 0.95\,$\pm$\,0.38 (15) \\
3--2              &    0.72\,$\pm$\,0.24 (13)       & 0.87\,$\pm$\,0.31 (10)  \\
\hline                                 
\end{tabular}
\tablefoot{Results are given as averages and standard deviations (including observational uncertainties). The number in parentheses gives the number of objects.}
\end{table}

%
%
\section{Summary}
\label{s:summary}

We report ALMA observations of the $^{12}$CO \mbox{1--0} (65), \mbox{2--1} (73), \mbox{3--2} (73), $^{13}$CO \mbox{2--1} (72), \mbox{3--2} (72) and C$^{18}$O \mbox{2--1} (14) lines, and continuum at 107 (3), 233 (18), and 338\,GHz (33), and APEX observations of $^{12}$CO  \mbox{2--1} (22), \mbox{3--2} (16), \mbox{4--3} (13), $^{13}$CO \mbox{2--1} (18), and \mbox{3--2} (13) lines towards a sample of 77 OH/IR stars within 3$^\circ$ of the Galactic centre (number of detected objects in parentheses; only three objects were not detected in any CO line). We complemented these data with infrared photometry and light curve analyses. In summary:\\

\noindent - Based on colour, CO line, and variability characteristics we introduced four categories of objects with different stellar and/or circumstellar characteristics: SdE/SgE/P (SSP, 40 objects), SdE/SgE/nP (SSnP, seven objects), DdE/SgE/nP (DSnP, ten objects), and DdE/CgE/nP (DCnP, 17 objects).

\noindent - For SgE objects, Moffat-functions were fitted to AARPs to estimate the sizes of the centrally peaked CO line brightness distributions. The angular extent at which the fit reaches 10\,\% of its peak value was adopted as a measure of the size. On average, for the $^{12}$CO lines, the extent of the 1$-$0 emission is $\approx$\,80\,\% larger than that of the 2$-$1 emission, which in turn is $\approx$\,60\,\% larger than the  3$-$2 emission. For $^{13}$CO, the 2$-$1 emission is about 65\,\% larger than the 3$-$2 emission, and, on average, the $^{12}$CO emissions are about 30\,\% larger than the corresponding $^{13}$CO emissions. Apart from differences in spatial extents, the morphology of the AARPs are similar for all objects.

\noindent - Line profiles were extracted from the centre pixels of four datasets for SgE objects: the original data cubes convolved so that they correspond to data observed with circular Gaussian beams of FWHMs of 0\farcs5, 1\arcsec, 2\arcsec, and 3\arcsec. In a few cases, we used apertures as small as 0\farcs2 to discriminate between circumstellar and surrounding CO line emission. For all lines, the 3\arcsec\ data contain at least 80\,\% of the total emission seen by the interferometer. 

\noindent - Results were obtained through line-profile fitting for SgE objects. The $^{12}$CO line shapes follow a trend of close to rectangular profiles for the 1$-$0 line to closer to parabolic profiles for the 3$-$2 line. The $^{13}$CO  line profiles are more flat-topped than the corresponding $^{12}$CO line profiles. The line-intensity ratios are: $^{12}$CO \mbox{2--1}/\mbox{1--0}\,=\,4.1, \mbox{3--2}/\mbox{2--1}\,=\,1.4, and \mbox{4--3}/\mbox{3--2}\,=\,1.0, and $^{13}$CO \mbox{3--2}/\mbox{2--1}\,=\,1.6, and $^{12}$CO/$^{13}$CO \mbox{2--1}\,=\,3.5 and \mbox{3--2}\,=\,3.1. There is only small scatter around the averages.

\noindent - The results suggest a homogeneous set of large-scale dust-CSE characteristics for  SSP and SSnP objects (Fig.~\ref{f:colour_colour}), and large-scale gas-CSE characteristics for SSP, SSnP, and DSnP objects (Tables~\ref{t:aarp_sizes}, \ref{t:int_ratio_sizes}, and \ref{t:int_ratios_transitions}).

\noindent - For CgE objects, the CO line brightness distributions are often extended and take the form of axi-symmetric structures in ten cases and  circular rings in two cases. For the rest, the sizes are too small to draw definitive conclusions. In terms of line shape, eight objects have extended line wings in addition to a central soft-parabola component. 

\noindent - The occurrence rate of the C$^{18}$O line is substantially higher for CgE than for SgE objects.  

\noindent - The gas expansion velocity distributions of the soft-parabola component appear very similar for all categories, the average is 15\,km\,s$^{-1}$. The full extents of the line wings of the CgE objects range from about 50 to 150\,km\,s$^{-1}$.

\noindent - In total three, 18, and 33 objects were detected in 107, 233, and 338\,GHz continuum, respectively. The detection rate is lowest for category SSP. The extent of the continuum emission is larger for CgE than SgE objects. The relative detection rates in the three bands are fully explained by a steep spectral index estimated to be $\approx$3.3 in the relevant frequency range. 

\noindent - Comparison between ALMA and APEX CO line flux densities indicate that the flux loss in the ALMA data is at the $\la$20\,\% level.

\noindent - In about 20\,\% of our objects we find surrounding structure in the CO line emission that we attribute to an interaction region between expanding circumstellar CO gas and the surrounding interstellar medium.

\noindent - The data presented here will be analysed in more  detail, and put into their contexts within the evolutionary scenario of AGB stars, in a series of forthcoming papers. The suspected interaction between circumstellar and interstellar gas will be discussed in a separate paper.

\section{Data availability}

Table C.1 is only available in electronic form at the CDS via anonymous ftp to {\tt cdsarc.u-strasbg.fr (130.79.128.5)} or via {\url{http://cdsweb.u-strasbg.fr/cgi-bin/qcat?J/A+A/}}.

\begin{acknowledgements}
This paper makes use of the following ALMA data: ADS/JAO.ALMA\#2023.1.00098.S. ALMA is a partnership of ESO (representing its member states), NSF (USA) and NINS (Japan), together with NRC (Canada) and NSC and ASIAA (Taiwan) and KASI (Republic of Korea), in cooperation with the Republic of Chile. The Joint ALMA Observatory is operated by ESO, AUI/NRAO and NAOJ. We acknowledge support from the Nordic ALMA Regional Centre (ARC) node based at Onsala Space Observatory. The Swedish Research Council grant No 2019-0020 supports the Nordic ARC node and the Swedish observations (in this case projects O-0111-9304 and O-0113-9300) on the Atacama Pathfinder EXperiment (APEX) telescope, operated by the Max Planck Institute for Radio Astronomy. This research has made use of the VizieR catalogue access tool, CDS, Strasbourg, France (the original description of the VizieR service was published in \citet{ochsetal00}), and the NASA/IPAC Infrared Science Archive (IRSA), which is funded by the US National Aeronautics and Space Administration (NASA) and operated by the California Institute of Technology. This publication makes use of data products from the Mid Course Space Experiment (MSX) operated by the  Ballistic Missile Defense Organization (BMDO), the Spitzer Space Telescope operated by the Jet Propulsion Laboratory, California Institute of Technology, under NASA contract 1407, the Visible and Infrared Survey Telescope for Astronomy (VISTA) operated by ESO, and the Wide-field Infrared Survey Explorer (WISE), a joint project of the University of California, Los Angeles, and the Jet Propulsion Laboratory/California Institute of Technology funded by NASA. We acknowledge the use of the ADS bibliographic services. TK and WV acknowledge support from the Olle Engkvist foundation through grant nr. 229-0368. NR acknowledges support from the Swedish Research Council (grant No. 2023-04744).
\end{acknowledgements}

\bibliographystyle{aa}
\bibliography{horef.bib}

\begin{appendix}

\onecolumn
\section{Sample}
\label{a:sample}

\FloatBarrier

\begin{longtable}{lccl}
\caption{\label{t:sample}Sample.} \\	
\hline \hline 
Object \,$^a$                                   &  $\alpha$(J2000)\,$^b$     & $\delta$(J2000)\,$^b$            &  Category \\
                                                        & [h:m:s]                               & [$^\circ$:\arcmin:\arcsec]   \\
\hline 
\endfirsthead
\caption{Continued.} \\
\hline \hline
Source\,$^a$                                   &  $\alpha$(J2000)\,$^b$     & $\delta$(J2000)\,$^b$            &  Category  \\
                                                        & [h:m:s]                               & [$^\circ$:\arcmin:\arcsec]     \\
\hline
\endhead
\hline  
\endfoot
\hline
\endlastfoot                                                       
OH357.092$-$0.362	                 & 17:39:57.49	               & --31:35:57.13     & DSnP \\
OH357.149$-$1.009	                 & 17:42:41.01	               & --31:53:37.50     & SSP \\
OH357.180$-$0.521	                 & 17:40:48.55	               & --31:36:34.73     & DSnP \\
OH357.474+0.367        	        & 17:38:01.59  	              & --30:53:17.66     & DCnP \\
OH357.638+1.890			& 17:32:29.63	               & --29:55:37.16     & SSP \\
OH357.675$-$0.060		        & 17:40:12.68	               & --30:56:43.42     & SSP \\
OH357.749+0.320	                 & 17:38:53.64	               & --30:40:48.45     & DSnP \\
OH357.819+1.990	                 & 17:32:33.87	               & --29:43:13.73     & SSP \\
OH357.980+0.826			& 17:37:28.98                 & --30:12:51.88     & DCnP \\
OH358.039$-$1.684	                 & 17:47:34.35                  & --31:29:09.85     & DCnP \\
OH358.052+1.304	                 & 17:35:48.21	               & --29:53:47.56     & SSP \\
OH358.083+0.137         	         & 17:40:26.49	               & --30:29:39.01     & DCnP \\
OH358.273$-$0.665	                 & 17:44:04.28	               & --30:45:22.08     & SSP \\
OH358.425$-$0.175	                 & 17:42:30.51	               & --30:22:07.83     & SSP \\
OH358.505+0.330			& 17:40:43.37                 & --30:02:04.72     & DCnP \\
OH358.522$-$1.061	                 & 17:46:14.96	               & --30:45:00.74     & SSP \\
OH358.720$-$0.620	                 & 17:44:58.91	               & --30:21:06.64     & SSP \\
OH359.011$-$0.116		        & 17:43:42.00	               & --29:50:21.94     & SSP \\
OH359.033+1.938	                 & 17:35:46.58	               & --28:43:45.25     & SSnP \\
OH359.117$-$0.169	                 & 17:44:09.85	               & --29:46:40.06     & SSP \\
OH359.140+1.137	                 & 17:39:07.70	               & --29:04:02.97     & DSnP \\
OH359.147+1.023	                 & 17:39:35.30	               & --29:07:24.43     & SSP \\
OH359.161$-$0.055		        & 17:43:49.53	               & --29:40:46.00     & SSP \\
OH359.201+0.285	                 & 17:42:35.49	               & --29:28:03.59     & SSnP \\
OH359.220+0.163			& 17:43:06.75	               & --29:30:56.94     & DCnP \\
OH359.233$-$1.876	                 & 17:51:12.13	               & --30:33:40.48     & DSnP \\
OH359.360+0.084			& 17:43:45.47	               & --29:26:17.41     & SSnP \\
OH359.380$-$1.201		         & 17:48:51.85	               & --30:05:18.96     & DCnP \\
OH359.467+1.029	                  & 17:40:20.33	               & --28:50:55.36     & SSP \\
OH359.486$-$2.942	                  & 17:56:04.40	               & --30:52:57.17     & SSP \\
OH359.500+2.776	                  & 17:33:43.27	               & --27:53:01.03     & SSP \\
OH359.543$-$1.775		          & 17:51:32.08                 & --30:14:36.33     & DCnP \\
OH359.564+1.287	                   & 17:39:34.80	               & --28:37:46.26     & SSP \\
OH359.581$-$0.240\,$^c$	          & 17:45:33.48	               & --29:25:07.38     & DCnP \\
OH359.632$-$0.431	                  & 17:46:25.94	               & --29:28:28.50     & SSP \\
OH359.664+0.636	                   & 17:42:20.47	               & --28:53:21.42     & SSP \\
OH359.731+1.260	                   & 17:40:05.30	               & --28:30:07.39     & SSP \\
OH359.745$-$0.404	                   & 17:46:35.66	               & --29:21:51.50     & SSP \\
OH359.750+2.629			  & 17:34:53.29                & --27:45:11.68     & DCnP \\
OH359.783$-$0.391	                   & 17:46:38.12	               & --29:19:30.96     & SSP \\
OH000.000+0.352			  & 17:44:14.95	               & --28:45:06.35     & SSP \\
OH000.024$-$0.874	                   & 17:49:06.23	               & --29:22:04.95     & SSP \\
OH000.071$-$0.205	                   & 17:46:35.32	               & --28:58:57.32     & SSP \\
OH000.072$-$2.044			 & 17:53:50.70                & --29:55:28.55     & DCnP \\
OH000.190+0.036	                   & 17:45:55.82	               & --28:45:18.84     & SSP \\
OH000.260+1.027			  & 17:42:15.58	               & --28:10:35.95     & DSnP \\
OH000.313+1.674			  & 17:39:53.90	               & --27:47:25.00     & SSP \\
OH000.319$-$0.041			  & 17:46:32.13	               & --28:41:04.49     & DSnP \\
OH000.333$-$0.181			 & 17:47:06.96                &	 --28:44:42.58     & DSnP \\
OH000.453$-$1.216	                  & 17:51:27.29	               & --29:10:32.72     & DSnP \\
OH000.484$-$0.167	                  & 17:47:25.05	               & --28:36:34.01     & SSP \\
OH000.517+0.050	                   & 17:46:39.03	               & --28:28:06.74     & SSP \\
OH000.621$-$0.661	                  & 17:49:40.20	               & --28:44:46.57     & SSP \\
OH000.667$-$0.035$^d$             &  17:47:20.19	               & --28:23:06.49     & -- \\
OH000.689+2.140                        &  17:39:00.95	               & --27:13:24.06     & SSP \\
OH000.729+0.451			  & 17:45:35.67	               & --28:04:45.30     & SSnP \\
OH000.810$-$1.959	                   & 17:55:13.28	               & --29:14:40.30     & SSP \\
OH000.814+0.179			  & 17:46:51.06	               & --28:08:52.20     & SSP \\
OH001.072+0.365			  & 17:46:44.32	               & --27:49:51.89     & DCnP \\
OH001.134$-$0.062$^d$             &  17:43:37.35	               & --27:13:08.89      & -- \\
OH001.184$-$0.958	                   & 17:52:08.82	               & --28:24:57.02     & SSP \\
OH001.221+0.294	                   & 17:47:21.74	               & --27:44:23.88     & SSP \\
OH001.227+2.005	                    & 17:40:48.77	               & --26:50:19.60     & SSP \\
OH001.234+1.273	                    & 17:43:37.34	               & --27:13:09.28     & SSnP \\
OH001.484$-$0.061	                    & 17:49:20.89	               & --27:41:54.37     & DSnP \\
OH001.794+2.078			   & 17:41:52.47	               & --26:19:11.57     & DCnP \\
OH001.803$-$0.047			  & 17:50:01.93               & --27:25:01.14     & DCnP \\
OH001.833$-$1.505	                   & 17:55:46.72	               & --28:07:56.80     & SSP \\
OH002.014$-$2.100	                   & 17:58:31.65	               & --28:16:24.42     & SSP \\
OH002.140$-$0.373	                   & 17:52:04.44	               & --27:17:39.51     & SSP \\
OH002.186$-$1.660	          	  & 17:57:11.43	               & --27:54:19.19     & DCnP \\
OH002.286$-$1.801		           & 17:57:58.27	               & --27:53:20.47     & DCnP \\
OH002.382+0.590	                    & 17:48:54.96	               & --26:35:34.45     & SSP \\
OH002.640$-$0.191			  & 17:52:30.79	               & --26:46:17.49     & DCnP \\
OH002.642+0.197$^d$                 &  17:51:01.48	              & --26:34:19.11     & -- \\
OH002.721$-$1.065	                   & 17:56:04.68	               & --27:08:35.53     & SSnP \\
OH002.726$-$0.352		           & 17:53:19.82                &	 --26:46:45.12     & SSnP \\
\hline					
\end{longtable}
\tablefoot{$^{(a)}$ Source names are taken from \citet{seveetal97a}. $^{(b)}$ Coordinates determined from $^{12}$CO(\mbox{3--2}) data. $^{(c)}$ Coordinates determined from 338\,GHz continuum. $^d$ No detection in CO nor in continuum, and coordinates taken from infrared data.}

\begin{figure}[h!]
\centering
  \includegraphics[width=8cm]{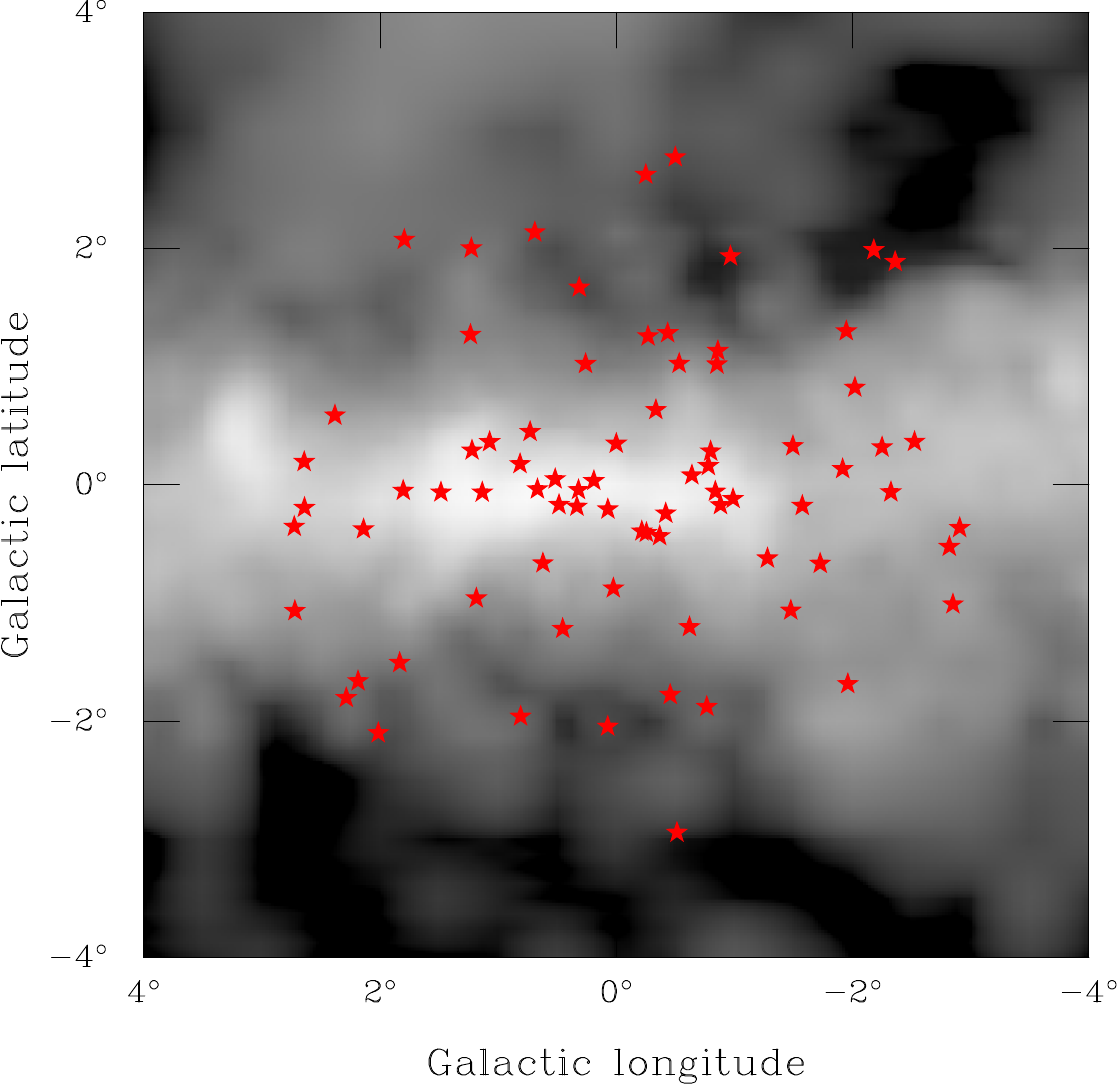}\\
    \caption{Positions of the sample objects marked as red stars on a $^{12}$CO($J$\,=\,1$-$0) map \citep{dameetal01}.} 
   \label{f:sample_co}
\end{figure}   

\begin{multicols}{2}
%
%
\section{Discussion of some individual objects}
\label{a:source_disc} 

Despite the filtering capacity of the interferometer, i.e. extended emission is resolved out and leaves no trace in the data, we have problems with contamination from non-circumstellar CO line emission in many cases. About 20\,\% of the objects show extended structures, often with low surface brightness, that can possibly be attributed to interaction between circumstellar and interstellar CO gas (CSM/ISM). Another about 10\,\% of the objects show clear presence of interstellar CO line emission in the velocity range of the circumstellar emission. As explained in Sect.~\ref{s:alma}, we refrained from using baseline-removal in most cases, and the data are handled in a way that minimizes the effects of any non-circumstellar emission as far as possible (see Sect.~\ref{s:line_prof_abc}). Below we comment on those objects where the presence of interstellar CO line emission, often in the form of CSM/ISM, complicates the extraction of reliable data.

\noindent OH357.474+0.367: A strong bow-shock-like feature $\approx$\,1\farcs5 from the object towards the south-west is most likely due to CSM/ISM. This affects the line intensities significantly when using increasingly larger apertures for extracting the spectra.

\noindent OH357.749+0.320: A weak arc-like feature $\approx$\,1\farcs2 from the object towards the east is most likely due to CSM/ISM. This affects the line intensities significantly when using increasingly larger apertures for extracting the spectra.

\noindent OH357.980+0.826: $^{12}$CO \mbox{1--0} and \mbox{2--1} affected by a narrow dip at the line centres. Most likely, extended CO line emission at this velocity leads to a significant loss of flux.

\noindent OH358.083+0.137: $^{12}$CO lines affected by weak, narrow dip at the line centres. Most likely, extended CO line emission at this velocity leads to a significant loss of flux.

\noindent OH358.273--0.665: A weak arc-like feature $\approx$\,1\farcs5 from the object towards the south is most likely due to CSM/ISM. This affects the line intensities when using increasingly larger apertures for extracting the spectra.

\noindent OH358.425--0.175: A weak arc-like feature $\approx$\,1\farcs7 from the object towards the north-west is most likely due to CSM/ISM. Strong, narrow dip at the centre of the $^{12}$CO line profiles ($^{13}$CO line data less affected). Most likely, extended CO line emission at this velocity leads to a significant loss of flux.

\noindent OH358.505+0.330: Strong and relatively broad dip at the centre of the $^{12}$CO line profiles. Most likely, extended CO line emission at this velocity leads to a significant loss of flux.

\noindent OH358.522--1.061: A weak arc-like feature $\approx$\,1\farcs5 from the object towards the south-east is most likely due to CSM/ISM. The line profiles are only weakly affected by this.

\noindent OH359.011--0.116: A strong bow-shock-like feature $\approx$\,0\farcs7 from the object towards the south-west is most likely due to CSM/ISM. This severely affects the line intensities when using increasingly larger apertures for extracting the spectra.

\noindent OH359.117--0.169: A weak arc-like feature $\approx$\,1\farcs5 from the object towards the south is most likely due to CSM/ISM. The line profiles are only weakly affected by this.

\noindent OH359.161--0.055: The $^{12}$CO data are severely affected by interstellar CO line emission, and 3\arcsec\ uvcut data are used in this case.

\noindent OH359.220+0.163: An arc-like feature $\approx$\,2\arcsec\ from the object towards the west is most likely due to CSM/ISM. The line profiles are strongly affected by this and interstellar CO line emission when using increasingly larger apertures for extracting the spectra.

\noindent OH359.360+0.084: A bow-shock-like feature $\approx$\,0\farcs7 from the object towards the south-west is most likely due to CSM/ISM. The line profiles, in particular those of $^{12}$CO, are severely affected by interstellar CO line emission. The $^{13}$CO line profiles are only weakly affected by this.

\noindent OH000.071--0.205: Features $\approx$\,1\arcsec\ -- 2\arcsec\ from the object towards the north-west may be due to CSM/ISM. The line profiles are affected by this when using increasingly larger apertures for extracting the spectra.

\noindent OH000.190+0.036: A strong bow-shock-like feature $\approx$\,0\farcs8 from the object towards the south-south-west is most likely due to CSM/ISM. This severely affects the line intensities when using increasingly larger apertures for extracting the spectra.

\noindent OH000.319--0.041: A weak shell-like feature $\approx$\,1\farcs6 from the object is most likely due to CSM/ISM. The $^{12}$CO line profiles are severely affected by this and additional interstellar CO line emission despite using the 3\arcsec\ uvcut data in this case. The $^{13}$CO line profiles are much less affected.

\noindent OH000.333--0.181: Several arc-like features within $\approx$\,1\farcs5 of the object are most likely due to CSM/ISM. This severely affects the line intensities when using increasingly larger apertures for extracting the spectra.

\noindent OH000.484--0.167: A weak feature $\approx$\,1\arcsec\ from the object towards the west may be due to CSM/ISM. The blue-shifted sides of the line profiles are affected by interstellar CO line emission.

\noindent OH000.517+0.050: An eye-like feature centred on the object ($\approx$\,4\arcsec$\times$3\arcsec ) is most likely due to CSM/ISM.  The line profiles are severely affected by this and additional interstellar CO line emission.

\noindent OH000.729+0.451: Possibly a weak bow-shock-like feature $\approx$\,0\farcs3 from the object towards the north-west can be due to CSM/ISM. The line profiles are strongly affected when using increasingly larger apertures for extracting the spectra.

\noindent OH000.814+0.179: The data are severely affected by interstellar CO line emission.

\noindent OH001.072+0.365: Possibly a weak bow-shock-like feature $\approx$\,2\farcs5 from the object towards the west that can be due to CSM/ISM. The line profiles are strongly affected when using increasingly larger apertures for extracting the spectra. 

\noindent OH001.803-0.047:  The $^{12}$CO data are affected by interstellar CO line emission.

\noindent OH001.221+0.294: A very weak shell-like feature $\approx$\,2\arcsec\ from the object is most likely due to CSM/ISM. The line profiles are strongly affected when using increasingly larger apertures for extracting the spectra.

\noindent OH002.140--0.373: The line profiles are affected when using increasingly larger apertures for extracting the spectra.

\noindent OH002.726--0.352: Features $\approx$\,0\farcs5 -- 1\farcs5 from the object towards the south-west may be due to CSM/ISM. This severely affects the line intensities when using increasingly larger apertures for extracting the spectra.

Finally, we comment on the three objects with no detections of circumstellar CO lines or continuum.

\noindent OH000.667-0.035: There is strong continuum and interstellar CO line emission in the vicinity of this object making it impossible to detect any circumstellar emission at the expected strengths.

\noindent OH001.134-0.062: There is interstellar CO line emission in the velocity range where the circumstellar CO line emission is expected to be, and there is no compact emission at the expected position of the object. This is somewhat surprising considering that the object falls in our category of large-amplitude, regular pulsators with a period of 500$^{\rm d}$ (see Sect.~\ref{s:variability}).

\noindent OH002.642+0.197: There is no apparent interstellar CO line emission in the vicinity of this object. There are weak $^{12}$CO \mbox{2--1} and \mbox{3--2} lines at +26\,km\,s$^{-1}$ with widths of about 16\,\,km\,s$^{-1}$. The centre velocity is slightly higher than the systemic velocity, +20\,km\,s$^{-1}$, and the width lower than that, 28\,km\,s$^{-1}$, expected from the OH 1612\,MHz data. The $^{13}$CO \mbox{2--1} and \mbox{3--2} lines show features more in accordance with the expected, a systemic velocity of about +19\,\,km\,s$^{-1}$ and an inferred expansion velocity of 15\,\,km\,s$^{-1}$. However, these lines are almost as strong as the corresponding possible $^{12}$CO lines, making a detection, at most, tentative. This is a large-amplitude pulsator with a long period, about 1520$^{\rm d}$ (see Sect.~\ref{s:variability}), so a detection of (strong) circumstellar CO line emission was expected.

\end{multicols}
\FloatBarrier
%
\section{ALMA CO line data}
\label{a:alma_line_intensities} 

\begin{landscape}
\begin{table*}[h!]
\caption{ALMA CO line and continuum data.}
\label{t:alma_line_continuum}
\begin{tabular}{lccccccccccccl} 
\hline \hline
\\[-2ex] 
Object                           &   $\varv_{\rm sys}$\,$^a$   & $\varv_\infty$\,$^a$    & $\varv_{\rm tot}$\,$^a$  &    & \multicolumn{4}{c}{$S$\,/\,$I$\,$^b$}                                                                                                                                                                                                                                 &   $\Theta_{0.1}$     &  \multicolumn{3}{c}{$S$\,$^c$}  &    Comments   \\ 
\cline{2-4} \cline{6-9}  \cline{11-13}
Line                      &  \multicolumn{3}{c}{[km\,s$^{-1}$]}                                                                    &     &  \multicolumn{4}{c}{[Jy]\,/\,[Jy\,km\,s$^{-1}$]}                                                                                                                                                                                                                               &      [\arcsec]       & \multicolumn{3}{c}{[mJy]}                 \\
                                                     
                             &                                           &                                    &      &                                  &        0\farcs5                                        &  1\farcs0                                               &  2\farcs0                                                                   &  3\farcs0                                                                       &                            &   107    &   233   &   338 \\
  \hline
 \\[-2ex]
OH357.092--0.362:  &   \phantom{1}72                 &     12                            & --   &                                 &                                                            &                                                              &                                                                                  &                                                                                    &                              &  --                 & --       & --   \\
$^{12}$CO(1--0)      &                                            &                                     &       &                                 &  0.029\,/\,0.54                                       &  0.049\,/\,0.99                                      &  0.069\,/\,\phantom{0}1.1\phantom{0}                      &  0.066\,/\,\phantom{0}1.1\phantom{0}                           &  0.61 \\
$^{12}$CO(2--1)     &                                             &                                     &      &                                  & 0.11\phantom{0}\,/\,2.1\phantom{0}     &  0.19\phantom{0}\,/\,3.3\phantom{0}    &  0.24\phantom{0}\,/\,\phantom{0}4.1\phantom{0}     &  0.24\phantom{0}\,/\,\phantom{0}4.2\phantom{0}        &   0.44 \\
$^{12}$CO(3--2)     &                                             &                                     &     &                                 &   0.22\phantom{0}\,/\,3.9\phantom{0}   &   0.31\phantom{0}\,/\,5.3\phantom{0}   &   0.33\phantom{0}\,/\,\phantom{0}5.9\phantom{0}    &   0.31\phantom{0}\,/\,\phantom{0}5.8\phantom{0}        &  0.36  \\
$^{13}$CO(2--1)     &                                             &                                     &     &                                  &   0.048\,/\,1.0\phantom{0}                    &   0.067\,/\,1.4\phantom{0}                     &   0.073\,/\,\phantom{0}1.8\phantom{0}                    &   0.083\,/\,\phantom{0}2.2\phantom{0}                           &  0.36\\
$^{13}$CO(3--2)    &                                             &                                     &      &                                 &  0.11\phantom{0}\,/\,1.9\phantom{0}     &  0.13\phantom{0}\,/\,2.3\phantom{0}    &  0.16\phantom{0}\,/\,\phantom{0}2.6\phantom{0}       &  0.16\phantom{0}\,/\,\phantom{0}2.7\phantom{0}       &  0.28  \\
C$^{18}$O(2--1)    &                                             &                                     &     &                                 &                                                               &                              &    --\,/\,--  \\
\\
OH357.149$-$1.009:  & 110                                  &     11                            &  --   &                             &                                                            &                                                              &                                                                                  &                                                                                    &                             &  --                 & --       & --     \\
$^{12}$CO(1--0)     &                                             &                                    &      &                               &  0.012\,/\,0.25                                       &  0.032\,/\,0.61                                       &  0.044\,/\,\phantom{0}0.81                                        &  0.039\,/\,\phantom{0}0.87                                            &  0.61 \\
$^{12}$CO(2--1)     &                                             &                                     &      &                                 & 0.075\,/\,1.3\phantom{0}                       &  0.14\phantom{0}\,/\,2.3\phantom{0}    &  0.19\phantom{0}\,/\,\phantom{0}2.8\phantom{0}     &  0.19\phantom{0}\,/\,\phantom{0}2.9\phantom{0}        &  0.41 \\
$^{12}$CO(3--2)     &                                             &                                     &      &                                 &  0.12\phantom{0}\,/\,2.1\phantom{0}    &   0.22\phantom{0}\,/\,3.8\phantom{0}   &   0.26\phantom{0}\,/\,\phantom{0}4.3\phantom{0}    &   --\,/\,--                                                                         &  0.35\\
$^{13}$CO(2--1)     &                                             &                                     &     &                                 &  0.026\,/\,0.60                                       &   0.049\,/\,0.84                                      &   0.068\,/\,\phantom{0}0.89                                       &   --\,/\,--                                                                         &  0.53\\
$^{13}$CO(3--2)    &                                             &                                     &     &                                 &  0.032\,/\,0.66                                       &  0.055\,/\,1.3\phantom{0}                      &  --\,/\,--                                                                       &  --\,/\,--                                                                           &  0.42\\
C$^{18}$O(2--1)    &                                             &                                     &     &                                 &                                                              &                                                               &   --\,/\,--  \\
\\
OH357.180--0.521: & \phantom{1}32                    &     13                            &  --  &                                 &                                                            &                                                              &                                                                                  &                                                                                    &                            &      --         &  \phantom{0}0.6             &  \phantom{0}3.2   \\
$^{12}$CO(1--0)     &                                             &                                    &       &                               &  0.029\,/\,0.38                                       &  0.044\,/\,\phantom{0}0.78                     &  0.068\,/\,\phantom{0}1.3\phantom{0}                     &  0.066\,/\,\phantom{0}1.6\phantom{0}                          &  0.59                    \\
$^{12}$CO(2--1)     &                                             &                                     &      &                                  & 0.11\phantom{0}\,/\,1.9\phantom{0}     &  0.18\phantom{0}\,/\,3.1\phantom{0}      &  0.23\phantom{0}\,/\,\phantom{0}4.0\phantom{0}    &  0.26\phantom{0}\,/\,\phantom{0}4.3\phantom{0}       &  0.36  \\
$^{12}$CO(3--2)     &                                             &                                     &     &                                 &   0.19\phantom{0}\,/\,3.4\phantom{0}   &   0.27\phantom{0}\,/\,4.7\phantom{0}     &   0.28\phantom{0}\,/\,\phantom{0}5.3\phantom{0}   &  0.25\phantom{0}\,/\,\phantom{0}5.0\phantom{0}        &  0.30  \\
$^{13}$CO(2--1)     &                                             &                                     &     &                                  &   0.049\,/\,1.0\phantom{0}                    &   0.065\,/\,1.3\phantom{0}                      &   0.068\,/\,\phantom{0}1.4\phantom{0}                     &   --\,/\,--                                                                        &  0.33 \\
$^{13}$CO(3--2)    &                                             &                                     &      &                                &  0.093\,/\,1.7\phantom{0}                      &  0.11\phantom{0}\,/\,2.0\phantom{0}      &  --\,/\,--                                                                      &  --\,/\,--                                                                         &   0.21 \\
C$^{18}$O(2--1)    &                                             &                                     &     &                                 &                                                               &                                                               &    --\,/\,--  \\
\\
OH357.474+0.367: & 117                                      &     17                            &  106 &                             &                                                            &                                                              &                                                                                  &                                                                                    &                       &     --      &  \phantom{0}1.9              & \phantom{0}5.9    & CO line CSM/ISM?  \\
$^{12}$CO(1--0)     &                                             &                                     &     &                           &  --\,/\,--                                                   &  --\,/\,--                                                    &  0.27\phantom{0}\,/\,\phantom{0}7.8\phantom{0}     &  --\,/\,--                                                                        &  --        \\
$^{12}$CO(2--1)     &                                             &                                     &     &                                  & --\,/\,--                                                  &  --\,/\,--                                                    &  1.1\phantom{00}\,/\,34\phantom{.00}                          &  --\,/\,--                                                                      &  -- \\
$^{12}$CO(3--2)     &                                             &                                     &     &                                  &   --\,/\,--                                                &  --\,/\,--                                                    &   1.8\phantom{00}\,/\,59\phantom{.00}                         &   --\,/\,--                                                                     &  -- \\
$^{13}$CO(2--1)     &                                             &                                     &     &                                  &   --\,/\,--                                                &  --\,/\,--                                                    &   0.40\phantom{0}\,/\,13\phantom{.00}                         &   --\,/\,--                                                                     &  -- \\
$^{13}$CO(3--2)    &                                             &                                     &      &                                &  --\,/\,--                                                  &  --\,/\,--                                                    &  0.72\phantom{0}\,/\,22\phantom{.00}                          &  --\,/\,--                                                                      &  -- \\
C$^{18}$O(2--1)    &                                             &                                     &      &                                &                                                              &                                                               &    0.049\,/\,\phantom{0}1.5\phantom{0}  \\
\hline
\end{tabular} 
\tablefoot{The full Table is available at the CDS. \,$^a$\,$\varv_{\rm sys}$ and $\varv_\infty$ are the averages of the  $^{12}$CO and $^{13}$CO 2$-$1 and 3$-$2 lines of the  0\farcs5, 1\farcs0, and 2\farcs0 datasets, while $\varv_{\rm tot}$ is the average FWZP of the $^{12}$CO and $^{13}$CO 2$-$1 and 3$-$2 lines for the DCnP objects. $^b$\,Flux density uncertainties are discussed in Sect.~\ref{s:line_prof_abc}. $^c$\,The continuum flux densities at the given frequencies in GHz.}
\end{table*}
\end{landscape}

\FloatBarrier
%

%
\section{APEX CO line data}
\label{a:apex_line_results}

\FloatBarrier
%
\begin{table*}[h!]
\centering
\caption{APEX CO line intensities.}
\label{t:apex_line_data}
 \begin{tabular}{lccccccl} 
\hline \hline
\\[-2ex] 
Object                              &    \multicolumn{6}{c}{$S$\,/\,$I$}                                                                                                   & Comment   \\ 
                                         &      \multicolumn{6}{c}{[Jy]\,/\,[Jy\,km\,s$^{-1}$]}                    \\
\cline{2-7}
\\[-2ex]
                                         &     \multicolumn{3}{c}{$^{12}$CO}                              &    &    \multicolumn{2}{c}{$^{13}$CO}   \\ 
\cline{2-4} \cline{6-7}
\\[-2ex]                                                       
                                       &  2--1                                                    &  3--2                                &  4--3                                                   &   &   2--1                                  &  3--2                              \\
  \hline
  \\[-2ex]
OH357.819+1.990	    &   1.5\phantom{0}\,/\,42\phantom{.0}    &   1.5\phantom{0}\,/\,33    &   1.9\phantom{0}\,/\,\phantom{1}50   &    &   $<$0.2\,/\,$<$4.8             &  0.40\,/\,12\phantom{.0}                     &           \\
OH358.052+1.304	    &   1.1\phantom{0}\,/\,21\phantom{.0}    &   1.3\phantom{0}\,/\,24    &   0.88\,/\,\phantom{1}18                     &    &   0.25\,/\,\phantom{1}6.3   &  $<$0.14\,/\,$<$4.3                            &            \\
OH359.033+1.938	    &   0.60\,/\,14\phantom{.0}                     &   0.58\,/\,14                      &   0.64\,/\,\phantom{1}12                     &    &   0.14\,/\,\phantom{1}3.3   &  0.22\,/\,\phantom{1}5.0                     &      \\
OH359.147+1.023	    &   1.2\phantom{0}\,/\,29\phantom{.0}    &   1.4\phantom{0}\,/\,37    &   1.0\phantom{0}\,/\,\phantom{1}28    &    &   0.14\,/\,\phantom{1}4.6   &  0.11\,/\,\phantom{1}4.7                     &       \\
OH359.380$-$1.201      &   1.4\phantom{0}\,/\,44\phantom{.0}    &   2.4\phantom{0}\,/\,88    &   3.2\phantom{0}\,/\,134                     &    &   0.35\,/\,14\phantom{.0}   &  0.90\,/\,32\phantom{.0}                     &   \\
OH359.486$-$2.942      &   0.49\,/\,11\phantom{.0}                      &   0.72\,/\,18                     &   0.49\,/\,\phantom{1}16                     &    &   0.14\,/\,\phantom{1}3.4   &  0.18\,/\,\phantom{1}3.2                     &     \\
OH359.500+2.776	    &   0.56\,/\,17\phantom{.0}                      &   0.83\,/\,20                     &   0.54\,/\,\phantom{1}18                     &    &   0.11\,/\,\phantom{1}3.3   &  $<$0.25\,/\,$<$3.2                             &     \\
OH359.750+2.629         &   2.84\,/\,53\phantom{.0}                      &   4.3\phantom{0}\,/\,82    &   7.4\phantom{0}\,/\,134                    &    &   0.74\,/\,17\phantom{.0}    &  1.4\phantom{0}\,/\,36\phantom{.0}   &    \\
OH359.783$-$0.391	    &   0.32\,/\,\phantom{0}6.7                      &   --                                   &   --                                                     &    &   0.05\,/\,\phantom{1}1.4    &  --                                                       &      \\
OH000.072$-$2.044	    &   0.39\,/\,\phantom{0}9.1                     &   --                                    &  --                                                      &    &   0.07\,/\,\phantom{1}2.0    &  --                                                       &   \\
OH000.313+1.674         &   0.39\,/\,\phantom{0}3.3                     &   --                                    &  --                                                      &    &   $<$0.05\,/\,$<$1.6           &  --                                                       &     \\
OH000.333$-$0.181      &   2.1\phantom{0}\,/\,34\phantom{.0}    &   6.1\phantom{0}\,/\,94     &   9.8\phantom{0}\,/\,150                   &    &   0.56\,/\,11\phantom{.0}     &  1.2\phantom{0}\,/\,23\phantom{.0}   &   ISM CO line contribution  \\
OH000.689+2.140         &   0.60\,/\,11\phantom{.0}                     &   0.76\,/\,17                       &   1.0\phantom{0}\,/\,\phantom{1}24  &    &   0.18\,/\,\phantom{1}3.9     &  0.25\,/\,\phantom{1}6.5                    &   \\
OH000.729+0.451         &   0.46\,/\,\phantom{0}4.6                     &   --                                    &   --                                                     &    &   $<$0.07\,/\,$<$2.1            &  --                                                      &    \\
OH000.810$-$1.959	    &   0.35\,/\,12\phantom{.0}                     &   0.79\,/\,22                      &   0.88\,/\,\phantom{1}22                    &    &   0.11\,/\,\phantom{1}2.7     &  $<$0.20\,/\,$<$3.2                            &     \\
OH001.184$-$0.958	    &   1.4\phantom{0}\,/\,17\phantom{.0}    &   2.6\phantom{0}\,/\,27    &   3.1\phantom{0}\,/\,\phantom{1}31   &    &   0.14\,/\,\phantom{1}3.1     &  0.58\,/\,\phantom{1}4.7                    &  ISM CO line contribution     \\
OH001.221+0.294	    &   2.6\phantom{0}\,/\,64\phantom{.0}    &   5.0\phantom{0}\,/\,90    &   --                                                     &    &   0.28\,/\,\phantom{1}8.1     &  0.50\,/\,\phantom{1}9.0                    &  ISM CO line contribution    \\
OH001.227+2.005	    &   0.67\,/\,10\phantom{.0}                     &   0.97\,/\,16                      &   --                                                     &    &   0.21\,/\,\phantom{1}4.2     &  0.29\,/\,\phantom{1}6.5                    &         \\
OH001.234+1.273         &   0.74\,/\,13\phantom{.0}                     &   0.86\,/\,16                      &   1.0\phantom{0}\,/\,\phantom{1}13   &    &   0.14\,/\,\phantom{1}3.5    &  0.11\,/\,\phantom{1}3.6                     &      \\
OH001.484$-$0.061	    &   0.53\,/\,11\phantom{.0}                     &   --                                    &   --                                                     &    &   $<$0.05\,/\,$<$1.6            &  --                                                      &         \\
OH001.794+2.078	    &   1.4\phantom{0}\,/\,36\phantom{.0}   &   2.2\phantom{0}\,/\,53     &   --                                                     &    &   0.32\,/\,11\phantom{.0}     &  0.61\,/\,16\phantom{.0}                    &    \\
OH002.286$-$1.801	    &   1.2\phantom{0}\,/\,30\phantom{.0}   &   --                                    &   --                                                     &    &   0.35\,/\,11\phantom{.0}     &  --                                                      &    \\
\hline
\end{tabular} 
\end{table*}

\section{Photometric variability results}
\label{a:variability_results}

\FloatBarrier

\setlength{\tabcolsep}{1.1mm}
\begin{longtable}{lrrccrrccrrcccc}
\caption{\label{t:variability_results}Results of the variability analysis.} \\
\hline
\hline
Object & $P_{\rm K}$ & $\sigma_{\rm P_K}$ &  $A_{\rm K}$   & $\sigma_{\rm A_K}$     & $P1$       & $\sigma_{\rm P1}$        &  $A1$       & $\sigma_{\rm A1}$      &  $P2$ & $\sigma_{\rm P2}$      & $A2$        & $\sigma_{\rm A2}$    & $P_{\rm adopted}$\,$^a$   & Variability\,$^b$ \\
                               &  [d]      &  [d]      &  [mag]     & [mag]    &     [d]      & [d]       &   [mag]   & [mag]    & [d]      & [d]     &   [mag]    & [mag)]   &  [d]     &type \\
\hline
\endfirsthead
\caption{Continued.}\\
\hline\hline
Object & $P_{\rm K}$ & $\sigma_{\rm P_K}$ &  $A_{\rm K}$   & $\sigma_{\rm A_K}$     & $P1$       & $\sigma_{\rm P1}$        &  $A1$       & $\sigma_{\rm A1}$      &  $P2$ & $\sigma_{\rm P2}$      & $A2$        & $\sigma_{\rm A2}$    & $P_{\rm adopted}$\,$^a$   &  Variability\,$^b$ \\
                               &  [d]      &  [d]      &  [mag]     & [mag]    &     [d]      & [d]       &   [mag]   & [mag]    & [d]      & [d]       &  [mag]   & [mag)]   &  [d]   & type \\
\hline
\endhead
\hline  
\endfoot
\hline
\endlastfoot  
OH357.092$-$0.362  &             &            &                &              &               &           &               &              &            &           &                &           &           & nP    \\
OH357.149$-$1.009  &  468     &    2.7   &  0.86       &  0.08     &   492      &   1.7   &  0.47      &  0.04     &   489   &   2.1   &  0.66       &  0.06  &   486  & LAP  \\
OH357.180$-$0.521  &             &            &                &              &               &           &               &              &            &           &                &           &           & nP      \\
OH357.474+0.367 &             &            &                &              &               &           &               &              &            &           &                &           &           & nP      \\
OH357.638+1.890  &  485    &    6.7   &  0.43       &  0.06     &   517      &   1.7   &  0.58      &  0.04     &   477   &   3.2   &  0.36       &  0.09   &   507  & LAP  \\
OH357.675$-$0.060  &  604     &   15.0  &  0.28       &  0.11     &   612      &   1.3   &  0.41      &  0.01     &   610   &   2.4   &  0.58       &  0.04   &   612  & LAP\\
OH357.749+0.320 &             &            &                &              &               &           &               &              &            &           &                &           &           & nP      \\
OH357.819+1.990  &  658    &    4.7   &  0.68       &  0.06     &   614      &   4.6   &  0.55      &  0.09     &   602   &   7.3   &  0.72       &  0.15  &   630  & LAP  \\
OH357.980+0.826  &             &            &                &              &               &           &               &              &            &           &                &           &           & nP      \\
OH358.039$-$1.684  &             &            &                &              &               &           &               &              &            &           &                &           &           & nP      \\
OH358.052+1.304  &  668    &   12.5  &  0.12       &  0.01     &   730      &   2.7   &  0.95      &  0.06     &   732   &   3.8   &  0.88       &  0.07  &   729  & LAP  \\
OH358.083+0.137  &            &            &                &              &               &           &               &              &            &           &                &            &           & nP     \\
OH358.273$-$0.665  &  530     &    2.0   &  0.74       &  0.03     &   523      &   2.7   &  0.50      &  0.02     &   522   &   0.8   &  0.69       &  0.02  &   523  & LAP  \\
OH358.425$-$0.175  &  1424   &  557    &  0.01       &  0.01     &  1867     &  32.3  &  0.54      &  0.10     &  1843  &  31.5  &  1.01       &  0.14  &   1854  & LAP  \\
OH358.505+0.330  &             &            &                &              &               &           &               &              &            &           &                &           &           & nP      \\
OH358.522$-$1.061  &  513     &    3.9    &  0.95      &  0.14     &   514      &   2.3    &  0.87     &  0.12      &   514   &   2.8   &  0.70       &  0.04  &   514  & LAP  \\
OH358.720$-$0.620  &  726     &   20.6   &  0.26      &  0.07     &   723      &   4.1    &  0.84     &  0.07      &   729   &   4.3   &  0.96       &  0.08  &   726  & LAP  \\
OH359.011$-$0.116   &             &            &               &               &   696     &   3.1    &  0.72     &  0.03      &   662    &   5.2   &  0.66      &  0.09  &   687  & LAP  \\
OH359.033+1.938  &             &            &                &              &               &           &               &              &            &           &                &           &           & nP      \\
OH359.117$-$0.169  &  963      &    5.7   &  1.15      &  0.07      &   965      &   4.7   &  0.84      &  0.07     &   983   &   5.7   &  0.85       &  0.06  &   970  & LAP  \\
OH359.140+1.137  &             &            &                &              &               &           &               &              &            &           &                &           &           & nP      \\
OH359.147+1.023  &  793     &   40.2  &  0.17       &  0.06     &   825      &   5.4   &  0.83      &  0.08     &   828   &    7.6  &  0.44       &  0.06  &   826  & LAP  \\
OH359.161$-$0.055  &  692      &    5.2   &  0.85       &  0.07     &   770      &   3.1   &  0.61      &  0.04     &   742   &   4.8   &  0.78       &  0.09  &   748  & LAP  \\
OH359.201+0.285  &             &            &                &              &               &           &               &              &            &           &                &           &           & nP      \\
OH359.220+0.163  &  677     &   82.8  &  0.01       &  0.03     &               &           &               &              &            &           &                &           &           & nP      \\
OH359.233$-$1.876  &             &            &                &              &               &           &               &              &            &           &                &        &           & nP \\
OH359.360+0.084  &            &            &                &              &               &           &               &               &   513   &   3.4  &  0.28       &  0.05   &           & nP \\
OH359.380$-$1.201  &             &            &                &              &               &           &               &              &            &           &                &           &           & nP      \\
OH359.467+1.029  &  605    &    2.8   &  1.03       &  0.04     &   603       &   1.4   &  0.89     &  0.06     &   607   &   3.4   &  0.73       &  0.06  &   604  & LAP  \\
OH359.486$-$2.942  &  689     &   11.4  &  0.71       &  0.02     &   632       &   2.5  &  1.12      &  0.05     &   634   &   7.1   &  0.96       &  0.07  &   635  & LAP  \\
OH359.500+2.776  &  852    &    8.5   &  1.15       &  0.07     &   835      &   6.3   &  1.01      &  0.13     &   827   &   5.0   &  0.76       &  0.08  &   834  & LAP  \\
OH359.543$-$1.775  &             &            &                &              &               &           &               &              &            &           &                &           &           & nP      \\
OH359.564+1.287  &  686    &    5.4   &  0.54       &  0.03     &               &           &               &              &   686   &   5.0   &  0.58       &  0.07  &   686  & LAP  \\
OH359.581$-$0.240  &             &            &                &              &               &           &               &              &            &           &                &           &           & ned      \\
OH359.632$-$0.431  &  636     &    3.2   &  0.91       &  0.01     &   654      &   3.1   &  0.83      &  0.07     &   650   &   2.2   &  0.86       &  0.03  &   648  & LAP  \\
OH359.664+0.636  &            &            &                &              &   535      &   3.4   &  0.55      &  0.06     &   575   &   4.1   &  0.41       &  0.06  &   551  & LAP  \\
OH359.731+1.260  &            &            &                &              &   481      &   1.6   &  0.29      &  0.01     &   473   &   2.4   &  0.39       &  0.07  &   479  & LAP  \\
OH359.745$-$0.404  &  733     &    3.8   &  0.79       &  0.01     &   748      &   5.5   &  0.63      &  0.06     &   749   &   6.1   &  0.75       &  0.09  &   740  & LAP  \\
OH359.750+2.629  &             &            &                &              &               &           &               &              &            &           &                &           &           & nP      \\
OH359.783$-$0.391  &  594     &    7.1    &  0.35       &  0.01     &   587      &   4.2   &  0.79     &  0.07     &   591   &   4.9    &  0.81      &  0.13  &   590  & LAP  \\
OH000.000+0.352  &            &             &                &             &   460       &   4.2   &  0.64     &  0.06     &   461   &   3.6    &  0.62      &  0.06  &   461  & LAP  \\
OH000.024$-$0.874  &  493     &    6.8    &  0.33       &  0.04    &   475      &   6.8    &  0.23     &  0.06     &   478   &   6.6    &  0.30      &  0.09  &   482: & LAP:  \\
OH000.071$-$0.205  &  718     &   36.5   &  0.09       &  0.03    &   800      &  11.2   &  0.48     &  0.09     &   781   &   3.0    &  0.86      &  0.07  &   782  & LAP  \\
OH000.072$-$2.044  &             &            &                &              &               &           &               &              &            &           &                &           &           & nP      \\
OH000.190+0.036  &  539    &    2.5   &  1.03       &  0.03     &   548      &   3.6   &  0.77      &  0.08     &   569   &   2.9    &  0.86      &  0.07  &   551  & LAP  \\
OH000.260+1.027  &             &            &                &              &               &           &               &              &            &           &                &           &           & nP      \\
OH000.313+1.674  &  487    &    8.8   &  0.39       &  0.15     &   478      &   1.7    &  0.68      &  0.05     &   474   &   2.8   &  0.49       &  0.08  &   477  & LAP  \\
OH000.319$-$0.041  &             &            &                &              &               &           &               &              &            &           &                &           &           & nP      \\
OH000.333$-$0.181  &             &            &                &              &               &           &               &              &            &           &                &           &           & nP      \\
OH000.453$-$1.216  &             &            &                &              &               &           &               &              &            &           &                &           &           & nP      \\
OH000.484$-$0.167  &             &            &                &              &   563      &   2.1   &  0.35      &  0.03     &   562   &   3.5   &  0.50       &  0.08  &   563  & LAP  \\
OH000.517+0.050  &            &            &                &              &               &           &               &              &   808   &   2.8   &  0.73       &  0.05  &   808  & LAP  \\
OH000.621$-$0.661  &  764     &    4.4   &  0.80       &  0.06     &   734      &   7.9   &  0.78      &  0.11     &   723   &  27.4  &  0.84       &  0.27  &   756  & LAP  \\
OH000.667$-$0.035  &             &            &                &              &               &           &               &              &            &           &                &           &           & nP      \\
OH000.689+2.140  &  550    &    2.4   &  0.87       &  0.03     &   561      &   3.2   &  0.71      &  0.05     &   562   &   1.6   &  0.59       &  0.03  &   559  & LAP  \\
OH000.729+0.451  &  645    &   46.0  &  0.03       &  0.01     &               &           &               &              &            &           &                &           &           & nP  \\
OH000.810$-$1.959  &  673     &    2.2   &  1.25       &  0.04     &   677      &   5.7   &  0.95      &  0.08     &   674   &   4.9   &  0.96       &  0.10  &   674  & LAP  \\
OH000.814+0.179  &  617    &    2.7   &  1.36       &  0.05     &   604      &   3.6   &  0.83      &  0.06     &   614   &   5.5   &  0.70        &  0.14  &   613  & LAP  \\
OH001.072+0.365  &             &            &                &              &               &           &               &              &            &           &                &           &           & nP      \\
OH001.134$-$0.062  &  507     &    4.5    &  0.53      &  0.07     &   496      &   1.9    &  0.31      &  0.02     &   500   &   0.8   &  0.55       &  0.01  &   500  & LAP  \\
OH001.184$-$0.958  &             &             &               &              &   488      &   2.0    &  0.88      &  0.07     &   492   &  15.5  &  0.40       &  0.35  &   488  & LAP  \\
OH001.221+0.294  &  699    &    2.5    &  1.29      &  0.04     &   713      &   3.2    &  0.68      &  0.08     &   739   &  23.5  &  0.28       &  0.18  &   705  & LAP  \\
OH001.227+2.005  &  551    &   28.8   &  0.11      &  0.03     &   572      &   5.4    &  1.10      &  0.09     &   571   &   1.8   &  0.73       &  0.06  &   571  & LAP  \\
OH001.234+1.273  &             &            &                &              &               &           &               &              &            &           &                &            &           & nP     \\
OH001.484$-$0.061  &             &            &                &              &               &           &               &              &            &           &                &           &           & nP      \\
OH001.794+2.078  &  441    &   16.5  &  0.05       &  0.04     &               &           &               &              &            &           &                &           &           & nP      \\
OH001.803$-$0.047  &             &            &                &              &               &           &               &              &            &           &                &           &           & nP      \\
OH001.833$-$1.505  &             &            &                &              &   600      &   2.5   &  0.74      &  0.07     &   597   &   3.7   &  0.72       &  0.09  &   599  & LAP  \\
OH002.014$-$2.100  &             &            &                &              &   355      &  11.2  &  0.41      &  0.12     &   336   &   3.9   &  0.53       &  0.20  &   338  & LAP  \\
OH002.140$-$0.373  &             &            &                &              &   565      &  10.8  &  0.92      &  0.12     &   557   &   5.6   &  0.73       &  0.12  &   559  & LAP  \\
OH002.186$-$1.660  &             &            &                &              &               &           &               &              &            &           &                &           &           & nP      \\
OH002.286$-$1.801  &             &            &                &              &               &           &               &              &            &           &                &           &           & nP      \\
OH002.382+0.590  &  542    &   28.2  &  0.22       &  0.13     &   543      &   8.1   &  0.51      &  0.11     &   545   &  10.6  &  0.52       &  0.19  &   544:  & LAP:\\
OH002.640$-$0.191  &             &            &                &              &               &           &               &              &            &           &                &           &           & nP\\
OH002.642+0.197  &  1423  &   19.5  &  1.57       &  0.11      &  1532    &   4.9   &  1.06      &  0.03     &  1510  &   9.7   &  1.04       &  0.05  &  1523  & LAP  \\
OH002.721$-$1.065  &             &            &                &              &               &           &               &              &            &           &                &           &           & nP      \\
OH002.726$-$0.352  &             &            &                &              &               &           &               &              &            &           &                &           &           & nP      \\
\hline
\end{longtable}
\tablefoot{Period with error, and amplitude with error in the $K_{\rm s}$ (2.15\,$\mu$m; columns 2, 3, 4, 6), $W1$ (3.4\,$\mu$m; columns 6, 7, 8, 9), and  $W2$ (4.6\,$\mu$m; columns 10, 11, 12, 13) bands. $^{(a)}$ The weighted average of the estimated periods. $^{(b)}$ LAP = large-amplitude, regular variability (see text for details), LAP: = regular variability not fully established, nP = enough data to conclude that the object has no regular variability, ned = not enough data.}

\end{appendix}

\end{document}